\documentclass[a4paper,UKenglish, autoref, thm-restate]{lipics-v2021}

\pdfoutput=1 
\hideLIPIcs  

\title{Menger Theorem for Temporal Paths (Not Walks)}

\author{Allen {Ibiapina}}{ParGO Group, Universidade Federal do Ceará, Brazil}{allen.ibiapina@alu.ufc.br}{https://orcid.org/0000-0002-6584-7718}{Funded by CAPES.}

\author{Raul {Lopes}}{{LAMSADE, Universit\'{e} Paris-Dauphine, PSL University, CNRS UMR7243, France}
\and {DIENS, Ecole normale supérieure de Paris, CNRS, France}}{rtlopes@protonmail.com}{https://orcid.org/0000-0002-7487-3475}
{group Casino/ENS Chair on Algorithmics and Machine Learning, and French National Research Agency under JCJC program (ASSK: ANR-18-CE40-0025-01).}

\author{Andrea {Marino}}{Universit\`{a} degli Studi di Firenze, Italy}{andrea.marino@unifi.it}{https://orcid.org/0000-0002-9854-7885}
{Italian PNRR CN4 Centro Nazionale per la Mobilità Sostenibile, NextGeneration EU - CUP,  B13C22001000001. MUR of Italy, under PRIN Project n. 2022ME9Z78 - NextGRAAL: Next-generation algorithms for constrained GRAph visuALization, PRIN PNRR Project n. P2022NZPJA - DLT-FRUIT: A user centered framework for facilitating DLTs FRUITion.}

\author{Ana {Silva}}
{ParGO Group, Universidade Federal do Ceará, Brazil}
{anasilva@mat.ufc.br}{https://orcid.org/0000-0001-8917-0564}
{CNPq Produtividade: 303803/2020-7,
CNPq Universal 404479/2023-5, and COFECUB 88887.712023/2022-00. This work was partially developed during this author's stay as visiting professor at the University of Florence.}

\authorrunning{A. Ibiapina, R. Lopes, A. Marino and A. Silva} 

\Copyright{Allen Ibiapina, Raul Lopes, Andrea Marino and Ana Silva} 



\category{} 

\relatedversion{} 

\nolinenumbers 

\EventEditors{John Q. Open and Joan R. Access}
\EventNoEds{2}
\EventLongTitle{42nd Conference on Very Important Topics (CVIT 2016)}
\EventShortTitle{CVIT 2016}
\EventAcronym{CVIT}
\EventYear{2016}
\EventDate{December 24--27, 2016}
\EventLocation{Little Whinging, United Kingdom}
\EventLogo{}
\SeriesVolume{42}
\ArticleNo{23}

\usepackage{url}
\usepackage{cite}

\usepackage{tikz}
\usetikzlibrary{positioning,fit,calc,shapes,backgrounds,snakes}
\usetikzlibrary{matrix}
\usetikzlibrary{arrows.meta}
\usetikzlibrary{calc}

\makeatletter
\newtheorem{repeatthm@}{Theorem}

\makeatother

\makeatletter
\newtheorem{repeatlmm@}{Lemma}

\makeatother

\makeatletter
\newtheorem{repeatpro@}{Proposition}

\makeatother

\usepackage{lipsum}
\usepackage{mwe}
\usepackage{hhline}

\usepackage{complexity}
\usepackage{environ,multirow}
\usepackage[ruled,vlined,english,linesnumbered]{algorithm2e}
\usepackage{setspace}

\usepackage{cancel}

\usepackage{xspace}
\usepackage{tabularx}

\makeatletter
\usepackage{environ}
\newcommand{\problemtitle}[1]{\gdef\@problemtitle{#1}}%
\newcommand{\probleminput}[1]{\gdef\@probleminput{#1}}%
\newcommand{\problemquestion}[1]{\gdef\@problemquestion{#1}}%
\NewEnviron{myproblem}{
  \problemtitle{}\probleminput{}\problemquestion{}%
  \BODY
  \par\addvspace{.5\baselineskip}
  \noindent
  \begin{tabularx}{\textwidth}{@{\hspace{\parindent}} l X c}
    \multicolumn{2}{@{\hspace{\parindent}}l}{\@problemtitle} \\
    \textbf{Input:} & \@probleminput \\
    \textbf{Question:} & \@problemquestion
  \end{tabularx}
  \par\addvspace{.5\baselineskip}
}

\usepackage{amsmath}
\usepackage{amsthm}
\usepackage{amsfonts}
\usepackage{amssymb}
\newtheorem{question}{Question}

\usepackage{soul}

\usepackage{soul} 

\definecolor{navyblue}{rgb}{0.0, 0.0, 0.5}
\newcommand{\change}[1]{#1} 

\definecolor{goodred}{HTML}{CC6677}
\definecolor{goodblue}{HTML}{332288}
\definecolor{goodyellow}{HTML}{DDCC77}
\definecolor{goodgreen}{HTML}{117733}
\definecolor{goodcyan}{HTML}{88CCEE}
\definecolor{goodwine}{HTML}{882255}
\definecolor{goodteal}{HTML}{44AA99}
\definecolor{goodolive}{HTML}{999933}
\definecolor{goodpurple}{HTML}{AA4499}

\usepackage{multirow}
\usepackage{xcolor,colortbl}

\newcommand{\tvertex}{temporal vertex\xspace}
\newcommand{\tvertices}{temporal vertices\xspace}

\newcommand{\twalk}[1]{$#1$-walk}

\newcommand{\tpath}[1]{$#1$-path}

\newcommand{\tvcut}[1]{\tvertex $#1$-cut}

\newcommand{\ktpaths}{$k$-\textsc{\tvertex-Disjoint paths}\xspace}
\newcommand{\htcuts}{$h$-\textsc{\tvertex path-Cut}\xspace}

\DeclareMathOperator{\pw}{\mathsf{w}} 
\DeclareMathOperator{\pwc}{\mathsf{wc}} 
\DeclareMathOperator{\tw}{\mathsf{tw}}
\DeclareMathOperator{\tc}{\mathsf{twc}} 
\DeclareMathOperator{\tp}{\mathsf{tp}} 
\DeclareMathOperator{\tpc}{\mathsf{tpc}} 

\newif\ifremoved
\removedtrue 

\usepackage{xcolor}
\newcommand{\lipItem}[1]{\textcolor{lipicsGray}{\sffamily\bfseries\upshape\mathversion{bold}#1}}
\definecolor{mid-green}{rgb}{0.15,0.65,0.15}
\definecolor{dark-blue}{rgb}{0.15,0.15,0.9}
\usepackage{hyperref}
\hypersetup{
  colorlinks, linkcolor={dark-blue},
citecolor={mid-green}, urlcolor={dark-blue}}
\usepackage[nameinlink]{cleveref} 
\crefname{proposition}{Prop.}{Props.} 
\Crefname{proposition}{Proposition}{Propositions}
\crefname{corollary}{Cor.}{Cors.}
\crefname{theorem}{Thm.}{Thms.}
\crefname{enumi}{Item}{Items}
\Crefname{enumi}{Item}{Items}

\begin{document}

\maketitle

\begin{abstract}
    A (directed) temporal graph is a (directed) graph whose edges are available only at specific times during its (discretized) lifetime $\tau$.
    In this setting, we ask that walks respect the temporal aspect by defining \emph{temporal walks} as sequences of adjacent edges whose appearing times are either strictly increasing or non-decreasing (here called non-strict), depending on the scenario.
    The notion of disjointness between walks is also not unique: two walks are \emph{vertex-disjoint} if they do not share a vertex, and are \emph{temporal vertex-disjoint} if they do not share a vertex at the same time.
    Thus a \emph{temporal path} is a temporal walk where no repetition of vertices, at any time, is allowed.
    This is an important distinction that separates the interpretation of our results from those of previous works on the topic.
    In this paper we focus on various questions regarding connectivity (maximum number of disjoint paths) and robustness (minimum size of a cut) between a given pair of vertices. Such problems are related to the well-known Menger's Theorem on static graphs. We explore all possible interpretations of such problems, according to vertex and temporal vertex-disjointness, strict and non-strict temporal paths, and directed and undirected temporal graphs. We present a number of new results, the main of which states that Menger's Theorem holds when the maximum number of \tvertex-disjoint temporal paths is equal to~1.
\end{abstract}

\setcounter{page}{1}

\section{Introduction}

Temporal graphs have been the subject of a lot of interest in recent years (see e.g. the surveys ~\cite{M.16,Netal.13,Holme.15,LVM.18}) and they have appeared under a variety of  names~\cite{Holme.15,B.96,XFJ.03,CFQS.12,LVM.18}.
Among the many applications (see e.g.~\cite{Holme.15}), consider Multi-Agent Path Finding problems (MAPF), an area that has grown considerably lately, and is often used by the AI and robotics communities (see e.g.~\cite{stern2019multi,stern2019multi2}). In it, a set of agents located at certain starting points of a network want to arrive at their targets by a sequence of actions (movements on the edges of the network) avoiding collisions (no two agents can meet at the same point at the same time). Translating to graph theory, it consists of finding vertex-disjoint walks.
As recently pointed out in~\cite{KlobasMMNZ21,KlobasMMNZ23,abs-2301-10503}, when we add time to the scenario, the situation can be seen as finding \tvertex-disjoint temporal walks. In other words, it is a generalization of a famous problem on static graphs, called the \textsc{$k$-Linkage} problem, known to be $\NP$-complete~\cite{karp1975computational}.
In this paper, we investigate the temporal version of a particular case of the \textsc{$k$-Linkage} problem; that in which all sources coincide, as well as all sinks. The latter one is known to be polynomial-time solvable on static graphs as it has ties with Menger's Theorem\cite{Menger.27}, as discussed in more detail later on.

Before we can start presenting our results, we need to introduce some crucial definitions.
We remark that, for convenience, all definitions and notations are include in~\autoref{sec:defs}.
A temporal graph is a graph\footnote{We use standard graph theory notation (see e.g.~\cite{West.book}).} where the edges are available only at prescribed moments.
More formally, for a integer $\tau$, a \emph{temporal graph with lifetime} $\tau$ is a pair ${\cal G} = (G,\lambda)$ where $G$ is a graph  and $\lambda$ is the \emph{time labeling} that assigns to each edge a finite subset of $[\tau]$, \change{where $[\tau]$ denotes the set $\{1,\ldots,\tau\}$}. Alternatively, a temporal graph can be seen as a finite sequence of spanning subgraphs of $G$ called \emph{snapshots}. A \emph{temporal vertex} (henceforth called \tvertex) is an occurrence of a vertex in time, i.e. an element of $V(G)\times[\tau]$, and a \emph{temporal edge} is an occurrence of an edge in time, i.e. $(e,i)$ with $e\in E(G)$ and $i\in \lambda(e)$.
Given a graph $G$ and vertices $s,t$, an \emph{$s,t$-walk} in $G$ is a sequence $(s=v_1,e_1,v_2,e_2\ldots,e_{p-1},v_p=t)$ of alternating vertices and edges of ${G}$ such that  $e_i=v_iv_{i+1}\in E({G})$, for every $i\in [p-2]$.
An \emph{$s,t$-path} is a walk that does not repeat vertices. Intuitively, a \emph{temporal $s,t$-walk (path)} is a walk (path) in $G$ that ``respects the flow of time''.
Formally, the edges within the walk or path are replaced by temporal edges, say $(e_1,i_1),\ldots,(e_{p-1},i_{p-1})$, such that $i_1\le \ldots \le i_{p-1}$. If instead we have only strict inequalities, we call such a walk or path \emph{strict} and say that it is \emph{non-strict}, otherwise.
We sometimes may write a temporal walk $(s = v_1, e_1, v_2, e_2, \ldots, v_{p-1}, v_p = t$) with $e_i = (v_iv_{i+1}, t_i)$, for $i < p$, as $(s=v_1,t_1,v_2,t_2\ldots,t_{p-1},v_p=t)$,
Observe that all these notions can easily be adapted to incorporate edge directions, meaning that we work on undirected and directed versions of the investigated problems. We always let it clear in the context which variation is being considered.
The strict model is more appropriate when one deals with fine granularity of time, while the non-strict one is more appropriate when the time granularity is relatively big, e.g. a single snapshot corresponds to all the streets available within a day (see~\cite{ZFMN.20}).

In classic graph theory, finding (internally) vertex-disjoint walks between a pair of vertices, $s$ and $z$, is equivalent to finding (internally) vertex-disjoint paths between $s$ and $z$. This is not the case for temporal graphs, as discussed later on. For now, let us go back to the MAPF problem previously defined.
Counterintuitively, it is not always the case that the robots prefer to take a path instead of a walk in order to arrive to its target. Walks might be preferred when an agent wants to make way for another agent to pass, which is the case for non-cooperative agents~\cite{GSLHW.22}.
Additionally, in the cooperative setting, paths might be preferred either because robots are autonomous systems and hence make their choices in a greedy way, going towards the destination and never coming back to a previous place (see e.g.~\cite{JMLW.22}), or in order to achieve computational efficiency by decreasing the size of the solution space (see e.g.~\cite{OD.22}).
This is why in ~\cite{KlobasMMNZ21,KlobasMMNZ23,abs-2301-10503} the authors propose to explore \tvertex-disjoint \emph{paths} between a given set of terminals.

Notice that, when removing the time labeling component (that is, considering static graphs), the MAPF problem previously described consists of, given a graph $G$ and pairs of vertices $\{(s_1,t_1),\ldots,(s_q,t_q)\}$, finding paths $P_1,\ldots,P_q$ that can intersect only in their endpoints and such that $P_i$ is an $s_i,t_i$-path , for every $i\in [q]$.
This is exactly the \textsc{$k$-Linkage} Problem that has already been mentioned and largely known to be $\NP$-complete~(see e.g.~\cite{karp1975computational}) within the Graph Theory community.
Therefore, adding time constraints to it can only make the problem harder. In this sense, the hardness results in~\cite{KlobasMMNZ21,KlobasMMNZ23,abs-2301-10503} are not so surprising as they generalize an already hard problem.
In this paper, we consider the generalization to temporal graphs of the easier problem, where all paths share the same origin and destination, which is known to be polynomial-time solvable on static graphs thanks to flow techniques and the famous Menger's Theorem~\cite{Menger.27}, as already mentioned.


\ifremoved{
    Our results then settle a gap in the knowledge by showing that, given a temporal graph and an integer $k$,  finding $k$ \tvertex-disjoint temporal paths is hard even with single origin and destination. If we are in directed temporal graphs, then the same holds even for fixed $k$ at least~3. On the other hand, we show that finding~2 such \tvertex-disjoint temporal paths is easy. This is a surprising result as all disjoint paths problems on temporal graphs investigated so far are either polynomial-time solvable for general $k$ or become hard for $k\ge 2$. As we will see, such result is made possible thanks to the fact that a Menger-like property holds when the maximum number of such paths is~1.
    Some of our results add to those results on static graphs that are known not to carry over to the temporal context (see~\cite{BF.03,B.96,KKK.00,CLMS.21,BM.23,MS.23}), and to the results that remark differences between temporal walks and paths (see~\cite{casteigts2021finding,FMNR.22} and previously cited~\cite{KlobasMMNZ21,KlobasMMNZ23,abs-2301-10503}).
}\fi

\paragraph*{Problem definition.}
The discussion that follows consider the non-strict case, but we make it clear from the context when the strict case is being applied.

We use \autoref{fig:walks_paths} for clarity purposes.
Observe that the temporal paths (not walks) from $s$ to $t$ are: $P_1=(s,1,u,2,t)$ $P_2=(s,1,u,3,t)$, $P_3=(s,2,u,2,t)$, and $P_4=(s,2,u,3,t)$. Temporal walks from $s$ to $t$ include all these temporal paths and the temporal walk $W_1=(s,1,u,1,x,2,y,3,u,3,t)$.
Finally, if we constrain ourselves to strict walks and paths, only $P_1$, $P_2$, and $P_4$ are picked.
Concerning disjointness, two possible versions can be considered. One asks agents to use different vertices, i.e. if an agent passes by a vertex $x$, then no other agent is allowed to pass by $x$, even at a different time. Another less constraining one asks that if an agent stays at vertex $x$ at time $i$, then no other agent can be at $x$ at time $i$, but could be at $x$ in another moment. The former is referred to as \emph{vertex-disjointness}, while the latter as \emph{\tvertex-disjointness}.
As an example of the difference between vertex-disjointness and \tvertex-disjointness, in \autoref{fig:walks_paths} note that $P_3$ and $W_1$ are not vertex-disjoint, as they both pass through $u$, but they are \tvertex-disjoint since they pass through $u$ at different times. As in~\cite{KlobasMMNZ21,KlobasMMNZ23,abs-2301-10503}, in this paper we deal with \tvertex-disjointness. More specifically, we consider the following two problems.

\begin{figure}
    \centering
    \begin{tikzpicture}[scale=0.7]
        \pgfsetlinewidth{1pt}
        \pgfdeclarelayer{bg}
        \pgfsetlayers{bg,main}

        \tikzset{vertex/.style={circle, minimum size=0.1cm, draw, inner sep=1pt, fill=black!20, ,thin}}
        \tikzset{snapshot/.style ={draw=black!50, rounded corners, dashed, minimum height=8mm, minimum width=2.3cm} }
        \tikzset{subgraph/.style ={draw=black!50, circle, draw, minimum width=1cm, yscale=2, fill=white} }

        \node[vertex] (s) at (-1,0) {$s$};
        \node[vertex] (u) at (2,0) {$u$};
        \node[vertex] (x) at (1,-2) {$x$};
        \node[vertex] (y) at (3,-2) {$y$};
        \node[vertex] (t) at (5,0) {$t$};

        \begin{pgfonlayer}{bg}
            \draw (s)--(u) node [midway,fill=white]{$\{1,2\}$}
            (u)--(x) node [midway,fill=white]{$\{1\}$}
            (x)--(y) node [midway,fill=white]{$\{2\}$}
            (y)--(u) node [midway,fill=white]{$\{3\}$}
            (u)--(t) node [midway,fill=white]{$\{2,3\}$};
        \end{pgfonlayer}

    \end{tikzpicture}
    \caption{The labelling $\lambda$ is represented on top of the edges. Temporal paths are: $P_1=(s,1,u,2,t)$ $P_2=(s,1,u,3,t)$, $P_3=(s,2,u,2,t)$, $P_4=(s,2,u,3,t)$. Additionally, $W_1=(s,1,u,1,x,2,y,3,u,3,t)$ is a temporal walk.
        \label{fig:walks_paths}}
\end{figure}

\begin{myproblem}
    \problemtitle{\textsc{(Directed)} \ktpaths}
    \probleminput{A temporal (directed) graph $(G,\lambda)$, vertices $s,t\in V(G)$, and an integer $k$.}
    \problemquestion{Are there at least $k$ (internally\footnote{We always consider paths that are internally disjoint and hence, from hereon, we omit the word ``internally''. }) \tvertex-disjoint temporal $s,t$-paths in $(G,\lambda)$?}
\end{myproblem}

Before we can state Menger's Theorem, consider the following definition and related problem. A \emph{\tvcut{s,t}} is a subset $S\subseteq (V(G)\setminus \{s,t\})\times [\tau]$ of temporal vertices intersecting every temporal $s,t$-path (not walk). In \autoref{fig:walks_paths}, $\{(u,1),(u,2)\}$ is a \tvcut{s,t} as it intersects $P_i$ for each $i\in [4]$. We emphasize the fact that $(s,1),\ldots,(s,\tau)$ and $(t_1,1),\ldots,(t_\tau,\tau)$ are not allowed in the cut.

\vspace{1mm}
\begin{myproblem}
    \problemtitle{\textsc{(Directed)} \htcuts}
    \probleminput{A temporal (directed) graph $\mathcal{G} = (G,\lambda)$, non-adjacent vertices $s,t\in V(G)$, and a positive integer $h$. }
    \problemquestion{Is there a \tvcut{s,t} of size at most $h$?}
\end{myproblem}

We also define the metrics in $(G,\lambda)$ related to the above problems:
\begin{itemize}
    \item $\tp_{\mathcal{G}}(s,t)$\ is the maximum integer $k$ such that the answer to \ktpaths\ is \textsc{yes}; and
    \item $\tpc_{\mathcal{G}}(s,t)$\ is the minimum integer $h$ such that the answer to \htcuts is \textsc{yes}.
\end{itemize}

In static graphs, the analogous parameters are equal by Menger's Theorem~\cite{Menger.27}. Formally, Menger's Theorem states that, given a graph $G$ and non-adjacent vertices $s,t\in V(G)$, the maximum number of vertex-disjoint $s,t$-paths in $G$ is equal to the minimum size of a vertex $s,t$-cut.
Asking whether Menger's Theorem holds also in the case of temporal (directed) graphs thus translates into asking whether $\tp_{\mathcal{G}}(s,t)=\tpc_{\mathcal{G}}(s,t)$ for every temporal graph and every pair of non-adjacent vertices $s,t$. As we will see shortly, the answer to such question is negative. 

%

Now, observe that we can interpret the temporal version of Menger's Theorem  in various ways, as we can consider either vertices or \tvertices, and consider either paths or walks. Below, we introduce the remaining possible interpretations and its respective metrics.

\begin{itemize}
    \item {$\pw_{\mathcal{G}}(s,t)$} is the maximum number of vertex-disjoint temporal $s,t$-walks and $\pwc_{\mathcal{G}}(s,t)$ is the minimum number of vertices intersecting every temporal $s,t$-walks. These values are equal if we consider paths instead of walks, as we show in \autoref{prop:vertex}. In \autoref{fig:walks_paths}, both values are equal to 1, as every walk passes by $u$.
    \item $\tw_{\mathcal{G}}(s,t)$ is the maximum number of \tvertex-disjoint temporal $s,t$-walks and $\tc_{\mathcal{G}}(s,t)$ is the minimum number of \tvertices intersecting every temporal $s,t$-walks. Note the difference from $\tp_{\mathcal{G}}(s,t)$ and $\tpc_{\mathcal{G}}(s,t)$, which refer to \emph{$s,t$-paths} rather than \emph{$s,t$-walks}. In \autoref{fig:walks_paths}, we have that $P_3$ and $W_1$ are temporal vertex-disjoint, while $\{(u,1),(u,2)\}$ form a \tvertex cut.
\end{itemize}
When it is clear from the context, we drop the ``$\mathcal{G}$'' from the subscript of all the aforementioned notations.

Note that we have not introduced metrics for the case of vertex-disjoint $\tpath{s,t}$s. This is because, as we prove in \autoref{prop:vertex}, this is the same solving the walks problems.

\paragraph*{Our results and related works.}
First, having investigated both the strict and non-strict cases, we must justify our earlier choice of notation. While we could have defined separate metrics for each case, further distinguishing between directed and undirected temporal graphs, doing so would have introduced excessive notation, making the manuscript unnecessarily cumbersome. For simplicity, we have chosen to define all metrics for undirected graphs in the non-strict case, making it clear from the context when directed and non-strict models are being considered. Now, we comment on \Cref{tab:summarywalks_nonstrict,tab:summarywalks_strict}, where we present our results. We first discuss the complexity results of each entry, then we talk about the various versions of Menger's Theorem.

\begin{table}[h]
    \renewcommand{\arraystretch}{1.2}
    \footnotesize
    \centering
        \begin{tabular}{|c|c|c|c|c|}
    \hline
    $X$ & $Y$ & \textsc{$\geq k$ $X$ Disjoint $Y$s} & Menger's Th.  & \textsc{$\leq h$ $X$ $Y$-Cut}\\
    \hline
    \hline
    \multirow{4}{*}{vertex} &       & \cellcolor{lightgray}$\pw(s,t):=\max_k$ & & \cellcolor{lightgray}$\pwc(s,t):=\min_h$ \\
    \cline{3-5}
                            & walk/path & \NP-c if $\tau=k\ge 2$   & \multirow{3}{*}{$\neq$ $\star$}  & \NP-c even if $\tau=2$ ,\\
    & (\cref{prop:vertex})  & for $G$ undirected $\star$  &  & $\W[1]$-hard  for $h$~\cite{ZFMN.20} \\

                            & & or directed ~\cite{B.96} &  & \\
     \hline
     \hline
     
    \multirow{8}{*}{t-vertex} & \multirow{2}{*}{walk} & \cellcolor{lightgray} $\tw(s,t):=\max_k$ & & \cellcolor{lightgray}$\tc(s,t):=\min_h$ \\
    \hhline{~~---}
    & & \cellcolor{red!10} Poly (\cref{thm:tvertex_vwalks})  & \cellcolor{red!10}$=$ (\cref{thm:tvertex_vwalks})  & \cellcolor{red!10}Poly (\cref{thm:tvertex_vwalks}) \\
    \hhline{~----}
    & \multirow{6}{*}{path} &  \cellcolor{lightgray}$\tp(s,t):=\max_k$ &  & \cellcolor{lightgray}$\tpc(s,t):=\min_h$ \\
    \hhline{~~---}
    & & \cellcolor{red!10}Poly if $k=1$ (\cref{prop:kequals1_easy}) & \cellcolor{red!10}$=$~(\cref{thm:menger}) & \cellcolor{red!10}Poly if $h=1$ (\cref{prop:kequals1_easy}) \\
    \hhline{~~---}
    &  & \cellcolor{red!10}Poly if $k=2$ (\cref{thm:finding-2-temporal-paths-poly-algorithm}) & \cellcolor{red!10}$\neq$ (\cref{prop:menger_counter})& \cellcolor{red!10}Poly if $h=2$ (\cref{cor:cuts_XP}) \\
    \hhline{~~---}
    &  & \cellcolor{red!10} \NP-c if $G$ undirected $\star$ & \cellcolor{red!10} & \cellcolor{red!10}co-\NP-hard $\star$\\
    &  & \cellcolor{red!10} or if $G$ directed, even if $\tau=3$ & \cellcolor{red!10}  & \cellcolor{red!10}$\XP$ for $h$ (\cref{cor:cuts_XP})\\
    &  & \cellcolor{red!10} for every fixed $k\ge 3$ (\cref{thm:paths_negative_directed}) & \multirow{-3}{*}{\cellcolor{red!10} $\neq$ (\cref{prop:menger_arbitrary_distance})} & \cellcolor{red!10} \\
    
    \hline
    \end{tabular}
    
    \caption{Results for the \textbf{non-strict} cases. Pink cells refer to our contributions. A $\star$ denotes a result that follows from the strict table by applying \autoref{thm:strict_to_nonstrict}. Shadow gray cells refer to the adopted notation, and ``t-vertex'' stands for ``\tvertex''. ``$\NP$-c'' stands for ``$\NP$-complete''. ``$\W[1]$-hard for $h$'' stands for ``$\W[1]$-hard when parameterized by $h$'' (similar for ``$\XP$ for $h$''). In column labeled ``Menger's Th.'', the sign ``$=$'' means that the related values coincide (i.e. Menger's Theorem holds), while ``$\neq$'' means they do not coincide.}
    \label{tab:summarywalks_nonstrict}
\end{table}

\begin{table}[h]
    \renewcommand{\arraystretch}{1.2}
    \footnotesize
    \centering
        \begin{tabular}{|c|c|c|c|c|}
    \hline
    $X$ & $Y$ & \textsc{$\geq k$ $X$ Disjoint $Y$s} & Menger's Th.  & \textsc{$\leq h$ $X$ $Y$-Cut}\\
    \hline
    \hline
    \multirow{4}{*}{vertex} &       & \cellcolor{lightgray}$\pw(s,t):=\max_k$ & & \cellcolor{lightgray}$\pwc(s,t):=\min_h$ \\
    \cline{3-5}
                            & walk/path & \NP-c if $G$ undirected   & \multirow{2}{*}{$\neq$ \cite{KKK.00}}  & \NP-c even if $\tau=5$,\\ 
   & (\cref{prop:vertex}) & even if &  & $\W[1]$-hard  for $h$~\cite{ZFMN.20} \\
                                    &         & $\tau=k\ge 2$~\cite{KKK.00} &  & \\
     \hline
     \hline
     
    \multirow{8}{*}{t-vertex} & \multirow{2}{*}{walk} & \cellcolor{lightgray} $\tw(s,t):=\max_k$ & & \cellcolor{lightgray}$\tc(s,t):=\min_h$ \\
    \hhline{~~---}
    & & \cellcolor{red!10} Poly \cite{MMS.19}  & \cellcolor{red!10}$=$ \cite{MMS.19}  & \cellcolor{red!10}Poly \cite{MMS.19} \\
    \hhline{~----}
    & \multirow{6}{*}{path} &  \cellcolor{lightgray}$\tp(s,t):=\max_k$ &  & \cellcolor{lightgray}$\tpc(s,t):=\min_h$ \\
    \hhline{~~---}
    & & \cellcolor{red!10}Poly if $k=1$ $\star$ & \cellcolor{red!10}$=$~$\star$ & \cellcolor{red!10}Poly if $h=1$ $\star$ \\
    \hhline{~~---}
    &  & \cellcolor{red!10}Poly if $k=2$ $\star$ & \cellcolor{red!10}$\neq$ $\star$ & \cellcolor{red!10}Poly if $h=2$ $\star$ \\
    \hhline{~~---}
    &  & \cellcolor{red!10} \NP-c if $G$ und. (\cref{thm:paths_negative_undirected});& \cellcolor{red!10} & \cellcolor{red!10}co-\NP-hard (item 2 of \cref{thm:decide_if_cut})\\
    &  & \cellcolor{red!10}  & \cellcolor{red!10}  & \cellcolor{red!10}$\XP$ for $h$ (\cref{cor:cuts_XP})\\
    &  & \cellcolor{red!10}  & \multirow{-3}{*}{\cellcolor{red!10} $\neq$ $\star$} & \cellcolor{red!10} \\
    
    \hline
    \end{tabular}
    
    \caption{Results for the \textbf{strict} cases. Pink cells refer to our contributions. A $\star$ denotes a result that follows from the strict table by applying \autoref{thm:strict_to_nonstrict}. Shadow gray cells refer to the adopted notation, and ``t-vertex'' stands for ``\tvertex''. ``$\NP$-c'' stands for ``$\NP$-complete''. ``$\W[1]$-hard for $h$'' stands for ``$\W[1]$-hard when parameterized by $h$'' (similar for ``$\XP$ for $h$''). In column labeled ``Menger's Th.'', the sign ``$=$'' means that the related values coincide (i.e. Menger's Theorem holds), while ``$\neq$'' means they do not coincide. }
    \label{tab:summarywalks_strict}
\end{table}

Concerning the first row of the tables ($X$ equal to ``vertex''), Berman~\cite{B.96} worked in the non-strict model and proved that deciding whether ${\cal G}$ has at least~$k$ vertex-disjoint temporal \tpath{s,t}s (or, equivalently, deciding whether $\pw(s,t)\ge k$ for a given $k$)  is $\NP$-complete when $G$ is undirected, and that it  remains hard when $k=\tau=2$ and $G$ is directed.
He then asked about the complexity when $k=2$ and ${G}$ is undirected. This was answered by Kempe et al.~\cite{KKK.00}, but only for the strict model.
This leaves open the complexity of deciding $\pw(s,t)\ge 2$ when $k=2$, $G$ is undirected, and the paths are non-strict.
Concerning vertex cuts (or equivalently, deciding whether $\pwc(s,t)\le h$), Zschoche et al.~\cite{ZFMN.20} prove that the problem is also $\NP$-complete even if $\tau=2$ for the non-strict case and if $\tau=5$ for the strict case.
They also prove that the problem is $\W[1]$-hard when parameterized by $h$ for both strict and non-strict contexts, and give an $\FPT$ algorithm when parameterized by $\tau+h$.  Finally, in~\cite{FMNRZ.20,MaackMNR23}, the authors investigate the vertex cut problem constrained to classes of temporal graphs.

As for row $X$ equal to ``\tvertex'' and $Y$ equal to ``walk'', we prove in \autoref{thm:tvertex_vwalks} that, in the non-strict case, all these values coincide (hence, the related version of Menger's Theorem holds) and they can be computed in polynomial time. As for the strict case, the analogous has already been proved in~\cite{MMS.19}. It is important also to mention that they use the nomenclature ``paths'' to refer to what are in fact ``walks''.
In \autoref{app:mertzios-paths-and-walks} we give a short discussion of why this is the case.


The situation is more interesting when $X$ is equal to ``vertex'' and $Y$ is equal to ``paths'', as can be seen in the tables.
At first glance the reader might think that the non-strict cases of the related problems are ``easier'' than the strict case. However, \autoref{thm:strict_to_nonstrict} proves the opposite: we provide a polynomial-time reduction from the (directed) strict case to the (directed) non-strict case. We mention additionally that the reduction preserves the natural parameters. This implies that any positive result for the non-strict case automatically extends to the strict case, while negative results for the strict case carry over to the non-strict case.
We use a star in the table to signal that the related entry is inherited from the other table based on \autoref{thm:strict_to_nonstrict}. 
We encourage the reader explore the tables and compare the complexity results between them. Next, we comment on column ``Menger's Th.''.

\begin{figure}
    \centering
    \begin{tikzpicture}[scale=0.7]
        \pgfsetlinewidth{1pt}
        \pgfdeclarelayer{bg}    
        \pgfsetlayers{bg,main}  

        \tikzset{vertex/.style={circle, minimum size=0.1cm, draw, inner sep=1pt, fill=black!20, ,thin}}
        \tikzset{snapshot/.style ={draw=black!50, rounded corners, dashed, minimum height=8mm, minimum width=2.3cm} }
        \tikzset{subgraph/.style ={draw=black!50, circle, draw, minimum width=1cm, yscale=2, fill=white} }

        \node[vertex] (u) at (0:0) {$u$};
        \node[vertex] (t) at (-22.5:3) {$t$};
        \node[vertex] (y) at (-67.5:3) {$y$};
        \node[vertex] (x) at (-112.5:3) {$x$};
        \node[vertex] (s) at (-157.5:3) {$s$};

        \begin{pgfonlayer}{bg}
            \draw (s)--(u) node [midway,fill=white]{$\{5\}$};
            \draw (s)--(x) node [midway,fill=white]{$\{1\}$};
            \draw (x)--(u) node [midway,fill=white]{$\{2\}$};
            \draw (x)--(y) node [midway,fill=white]{$\{4\}$};
            \draw (y)--(u) node [midway,fill=white]{$\{6\}$};
            \draw (y)--(t) node [midway,fill=white]{$\{7\}$};
            \draw (u)--(t) node [midway,fill=white]{$\{3\}$};
        \end{pgfonlayer}

        \begin{scope}[xshift=9cm]
            \node[vertex] (u) at (0:0) {$u$};
            \node[vertex] (t) at (-22.5:3) {$t$};
            \node[vertex] (y) at (-67.5:3) {$y$};
            \node[vertex] (x) at (-112.5:3) {$x$};
            \node[vertex] (s) at (-157.5:3) {$s$};

            \begin{pgfonlayer}{bg}
                \draw[->] (s)--(u) node [midway,fill=white]{$\{5\}$};
                \draw[->] (s)--(x) node [midway,fill=white]{$\{1\}$};
                \draw[->] (x)--(u) node [midway,fill=white]{$\{2\}$};
                \draw[->] (x)--(y) node [midway,fill=white]{$\{4\}$};
                \draw[->] (y)--(u) node [midway,fill=white]{$\{6\}$};
                \draw[->] (y)--(t) node [midway,fill=white]{$\{7\}$};
                \draw[->] (u)--(t) node [midway,fill=white]{$\{3\}$};
            \end{pgfonlayer}
        \end{scope}

    \end{tikzpicture}
    \caption{Example given in~\cite{KKK.00}, where $\pw(s,t)=1<\pwc(s,t)=2$.
        Observe that all the \tpath{s,t}s passing by $u$ uses either $x$ or $y$. This leads to the conclusion that there are no 2~vertex-disjoint temporal $s,t$-paths. Additionally, none of $u,x,y$ breaks all \tpath{s,t}s by itself. Therefore, the minimum size of an $s,t$-cut is equal to~2.}
    \label{fig:KKKexample}
\end{figure}

In the first row ($X$ equal to ``vertex''), the example to the left of \autoref{fig:KKKexample} tells us that, in the vertex-disjoint versions of the problems, Menger does not hold. Observe that all paths are strict, hence the example also works for the non-strict context. Additionally, observe that the example to the right closes all the gaps within this cell of the table, as it shows that the same is true on directed temporal graphs. We encourage the reader to see that the trick used in \autoref{prop:menger_arbitrary_distance} can be used to produce graphs where the difference between the size of a minimum cut and the maximum number of vertex-disjoint paths can be made arbitrarily large.

As we have already seen that Menger's Theorem works on \tvertex-disjoint walks, we jump directly to its behavior within the context of \tvertex-disjoint paths, where we prove that it holds only when the maximum number of temporal paths is equal to 1. This is the main result of our paper, and is presented in \autoref{sec:paths_algorithm}, \autoref{thm:menger}. Before we move on, we give an example where the number of disjoint paths is~2, while one needs to remove 3 \tvertices in order to break all the paths. Observe \autoref{fig:tvertexpaths_counter}. We argue in the preliminaries that indeed this is the case (\autoref{prop:menger_counter}). In fact, \autoref{prop:menger_arbitrary_distance} gives us that the difference between the parameters can be made arbitrarily large. Additionally, the construction in \autoref{thm:strict_to_nonstrict} allows us to get also a counter-example for the problem with the same metric values, but with larger lifetime (namely, lifetime 6). Finally, we bring attention to the fact that the example given in the figure is best possible. In other words, by \autoref{thm:menger}, there cannot be an example where $\tp(s,t)=1 < \tpc(s,t)=2$.

Finally, an unexpected byproduct of our results (specifically, of \autoref{thm:menger} and \autoref{cor:cuts_XP}) is that we can find 2 \tvertex-disjoint \tpath{s,t}s, if they exist, in polynomial time even in the undirected and non-strict context. This might suggest that {\ktpaths} is $\XP$ when parameterized by $k$ for $G$ undirected, which would  be in stark constrast with the case of $G$ directed, as we have proved that already for $k=\tau=3$ the problem is $\NP$-complete (\autoref{thm:paths_negative_directed}).

\begin{figure}
    \begin{center}
        \begin{tikzpicture}[scale=0.8]
  \pgfsetlinewidth{1pt}
  \pgfdeclarelayer{bg}    
   \pgfsetlayers{bg,main}  

  \tikzset{vertex/.style={circle, minimum size=0.1cm, draw, inner sep=1pt, fill=black!20, ,thin}}
  \tikzset{snapshot/.style ={draw=black!50, rounded corners, dashed, minimum height=8mm, minimum width=2.3cm} }
  \tikzset{subgraph/.style ={draw=black!50, circle, draw, minimum width=1cm, yscale=2, fill=white} }
  
    \node[vertex] (s) at (-3,0) {$s$};
    \node[vertex] (u) at (0,0) {$u$};
    \node[vertex] (x) at (0,2) {$x$};
    \node[vertex] (y) at (0,-2) {$y$};
    \node[vertex] (t) at (3,0) {$t$};
    
    \begin{pgfonlayer}{bg}    
        \draw (s) edge [out=90,in=180] node [midway,fill=white]{$\{1,2\}$} (x);
        \draw (s) edge [out=-90,in=180] node [midway,fill=white]{$\{1\}$} (y);
        \draw (x) edge [out=0,in=90] node [midway,fill=white]{$\{2,3\}$} (t);
        \draw (y) edge [out=0,in=-90] node [midway,fill=white]{$\{2,3\}$} (t);
        \draw (x) edge  node [midway,fill=white]{$\{1,2,3\}$} (u);
        \draw (u) edge  node [midway,fill=white]{$\{1,2\}$} (y);
    \end{pgfonlayer}

    \begin{scope}[xshift=9cm]
      \node[vertex] (s) at (-3,0) {$s$};
      \node[vertex] (u) at (0,0) {$u$};
      \node[vertex] (x) at (0,2) {$x$};
      \node[vertex] (y) at (0,-2) {$y$};
      \node[vertex] (t) at (3,0) {$t$};
    
      \begin{pgfonlayer}{bg}    
        \draw[->] (s) edge [out=90,in=180] node [midway,fill=white]{$\{1,2\}$} (x);
        \draw[->] (s) edge [out=-90,in=180] node [midway,fill=white]{$\{1\}$} (y);
        \draw[->] (x) edge [out=0,in=90] node [midway,fill=white]{$\{2,3\}$} (t);
        \draw[->] (y) edge [out=0,in=-90] node [midway,fill=white]{$\{2,3\}$} (t);
        \draw[<->] (x) edge  node [midway,fill=white]{$\{1,2,3\}$} (u);
        \draw[<->] (u) edge  node [midway,fill=white]{$\{1,2\}$} (y);
      \end{pgfonlayer}
    \end{scope}

  \end{tikzpicture}
    \end{center}
    \caption{Example where $\tp(s,t)=2<\tpc(s,t)=3$ in the \emph{non-strict} context. This is proved in \autoref{prop:menger_counter}. In fact, \autoref{prop:menger_arbitrary_distance} tells us that the difference $\tpc(s,t) - \tp(s,t)$ can be arbitrarily large.}
    \label{fig:tvertexpaths_counter}
\end{figure}

\paragraph*{Further Related Work.}
\label{sec:related}
A concept on static graphs that has an important impact on the study of temporal paths and walks is that of bounded length paths. In it, given a graph $G$, vertices $s$ and $t$, and an integer $b$, one is interested in maximizing the number of $s,t$-paths of length at most $b$ that are internally vertex-disjoint, and on minimizing the number of vertices needed to break all such paths. These concepts have been largely investigated and a complete picture can be found in~\cite{GT.11}. Observe that, if we consider only strict temporal paths, then such paths on $G$ are equivalent to vertex-disjoint strict temporal $s,t$-paths in the temporal graph $(G,\lambda)$ where $\lambda(e)$ is equal to $[b]$ for every $e\in E(G)$.
This immediately gives many of the cited complexity results presented in~\cite{MMS.19}. An interesting aspect of these paths on static graphs is that a version of Menger's Theorem holds in case $b\le 4$ (Lovász, Neumann-Lara and Plummer~\cite{LNP.78}). Observe that this does not translate directly to vertex-disjoint strict temporal paths, as we do not have a reduction in the opposite direction. Nevertheless, our version of Menger's Theorem might be seen as the first one on temporal paths similar to the previously cited one~\cite{LNP.78}.

\paragraph*{Structure of the paper.} \autoref{sec:defs} introduces all the necessary definitions and notation, as well as present some basic properties and results of the investigated problems. In \autoref{sec:negative}, we present our negative results and in \autoref{sec:positive_results}, we present our positive results. Finally, we conclude the paper with some open questions.


\section{Definitions and preliminary results}\label{sec:defs}

Given a natural number $n$ we denote by $[n]$ the set $\{1,2,\ldots,n\}$. Our definitions for graphs are standard and we refer the reader to~\cite{bang2008digraphs,West.book}. We also refer the reader to~\cite{CyganFKLMPPS15} for the parameterized complexity notation. 

A \emph{temporal (directed) graph} with \emph{lifetime} $\tau$ is a pair ${\cal G} = (G,\lambda)$ where $G$ is a (directed) graph  and
$\lambda$ is a function that assigns to each edge a subset of $[\tau]$.
We say that $G$ is the \emph{underlying graph} of ${\cal G}$.
We call an integer $p \in [\tau]$ a \emph{timestep}, an element $(v,t)\in V(G)\times [\tau]$ a \emph{temporal vertex}, and an element $(e,t)\in E(G)\times [\tau]$ such that $t\in \lambda(e)$ a \emph{temporal edge}.
We also sometimes call a temporal vertex $(u,i)$ a \emph{copy of $u$}.
Given $e\in E(G)$ and $i\in \lambda(e)$, we say that $e$ is \emph{active at timestep $i$}.
Also, we call the subgraph of $G$ formed by the edges active at timestep $i\in[\tau]$ the \emph{$i$-th snapshot of ${\cal G}$}, and we denote it by $G_i$.
Additionally, we use $V({\cal G}),E({\cal G})$ to denote $V(G), E(G)$, respectively.
We denote the set of temporal vertices of ${\cal G}$ by $V^T({\cal G})$, and the set of temporal edges by $E^T({\cal G})$.

Given a temporal (directed) graph ${\cal G}=(G,\lambda)$ and $s,t \in V(G)$, a \emph{temporal \twalk{s,t}} is a sequence $(s=v_1,e_1,v_2,e_2\ldots,e_{p-1},v_p=t)$ of alternating vertices of $G$ and temporal edges of ${\cal G}$ such that  $e_i=(v_iv_{i+1},t_i)\in E^T({\cal G})$ for each $i$ with $i<p$,
and $t_1\le \ldots\le t_{p-1}$.
It is called \emph{strict} if $t_1<\ldots < t_{p-1}$; and we sometimes write simply $(s=v_1,t_1,v_2,t_2\ldots,t_{p-1},v_p=t)$.
If $v_1,\ldots,v_p$ are all distinct, then we say that $P$ is a \emph{temporal \tpath{s,t}}.
Let $\gamma$ be $0$ for the non-strict case and $1$ for the strict one.
Then, for both cases, for each $i\in\{2,\ldots,p-1\}$, we say that the walk/path \emph{waits in $v_i$ from timestep $t_i+\gamma$ to $t_{i+1}$}, and we say that $P$ \emph{contains} a temporal vertex $(v_i,j)$ if either $v_i=s$ and $j=t_1$, or $v_i=t$ and $j=t_{p-1}+\gamma$, or $P$ waits on $v_i$ at time $j$.

We emphasize the fact that we assume that the endpoints occur only once in $P$. We also say that $P$ \emph{contains} the edge $v_iv_{i+1}$ and the temporal edge $e_i$, for each $i\in [p-1]$.
Finally, we denote the set of vertices of $P$ by $V(P)$, and the set of temporal vertices contained in $P$ by $V^T(P)$.
Similarly, we denote by $E(P)$ the set of edges contained in $P$ and by $E^T(P)$, the set of temporal edges contained in $P$. From now on we refer to temporal walks/paths simply as walks/paths, and we make it clear when we are dealing with a walk/path in the underlying graph.

Given two \twalk{s,t}s $P$ and $Q$, we say that $P$ and $Q$ are \emph{vertex-disjoint} if they do not pass through a same vertex, i.e. $V(P)\cap V(Q)=\{s,t\}$, while we say they are \emph{\tvertex-disjoint} if they do not pass through a same temporal vertex, i.e. $V^T(P)\cap V^T(Q)\subseteq \{s,t\}\times[\tau]$. The latter condition means that $P$ and $Q$ can pass through a same vertex but at different times.

The next proposition basically tell us that, when considering vertex-disjointness, the problems for walks and for paths collapse. This is why we define only two parameters for this variation, namely $\pw(s,t)$ and $\pwc(s,t)$.

\begin{proposition}
    Let $(G,\lambda)$ be a temporal (directed) graph and $s,t\in V(G)$. Then, the maximum number of vertex-disjoint \tpath{s,t}s is equal to the maximum number of vertex-disjoint \twalk{s,t}s. Similarly, the minimum size of $S\subseteq V(G)$ such that every \tpath{s,t} contains some $x\in S$ is equal to the minimum size of $S'\subseteq V(G)$ such that every \twalk{s,t} contains some $y\in S'$. These statements hold for the strict and the non-strict models.
    \label{prop:vertex}
\end{proposition}
\begin{proof}
    We let the reader observe that the arguments apply for directed and undirected graphs, and in the non-strict and the strict contexts.

    We first prove the first part of the statement, namely the one concerning paths and walks. Let $W$ be an \twalk{s,t} such that $W$ uses a vertex $u$ at least twice. More precisely: $W_{1}=(s,\ldots,i_1,u,i_2, \ldots, h_1,u,h_2,\ldots,t).$ Then, we can obtain a temporal walk following $W_1$ until timestep $i_{1}$ and, instead of using the temporal edge incident to $u$ at timestep $i_{2}$ to leave $u$, we use the temporal edge appearing at timestep $h_2$ and finish the temporal walk as done in $W$. Calling this temporal walk $W'$ we have that $W'$ has less vertex repetitions than $W$ and $V(W')\subseteq V(W)$. By repeating this process we obtain an \tpath{s,t} $P$ such that $V(P) \subseteq V(W)$. Now let $W_{1},\ldots,W_{k}$ be a set of vertex-disjoint \twalk{s,t}s, and consider $P_1,\cdots,P_k$ \tpath{s,t}s, where each $P_i$ is obtained from $W_i$ as explained. Then $P_{1}, P_{2},\ldots, P_{k}$ is a set of vertex-disjoint \tpath{s,t}s. Indeed if $V(W_i)\cap V(W_j)=\emptyset$ for every $i,j\in [k]$ with $i\neq j$, since $V(P_{i})\subseteq V(W_i)$ for every $i\in [k]$, we get that $V(P_{i})\cap V(P_{j})=\emptyset$ for every $i,j\in [k]$ with $i\neq j$, as desired.

    For the second part of the statement, observe that the paragraph above also gives us a way to obtain a \twalk{s,t} starting from a \tpath{s,t}. Since every \tpath{s,t} is also a \twalk{s,t}, note that the proposition follows.
\end{proof}

Given temporal (directed) graphs ${\cal G} = (G,\lambda)$ and ${\cal H} = (H,\lambda')$, we say that ${\cal H}$ is a \emph{temporal subgraph of ${\cal G}$} if $H\subseteq G$ and $\lambda'(e)\subseteq \lambda(e)$ for every $e\in E(H)$; we write ${\cal H}\subseteq {\cal G}$.

Given a temporal edge $(xy,i)$ of ${\cal G}$, we denote by ${\cal G}-(xy,i)$ the temporal graph obtained from ${\cal G}$ by removing $i$ from $\lambda(xy)$.
Given an \twalk{x_1,x_{q+1}}, $W=(x_1,e_1\ldots,e_q,x_{q+1})$, and $i<j$ we denote by $x_iPx_j$ the \twalk{x_i,x_j} contained in $W$.  Also, if $x_i=x_j=x$ for $i<j$, let $e_i = (x_ix_{i+1},h_1)$ and $e_{j-1} = (x_{j-1}x_j,h_2)$ (i.e., $h_1$ is the time of departure from $x$ through $e_i$ and $h_2$, the time of return to $x$ through $e_{j-1}$).
We denote the \twalk{x_1,x_{q+1}} obtained from $W$ by waiting on $x$ from time $h_1$ to time $h_2$ by $x_1W(x,h_1)..(x,h_2)Px_{q+1}$.
We sometimes also use $(x_i,h_1)Wx_j$ to denote the \twalk{x_i,x_j} contained in $W$ in order to emphasize the time of departure from $x_i$; similarly we might use $x_iW(x_j,h_2)$ or even $(x_i,h_1)W(x_j,h_2)$.
Additionally, if $Q$ is an \twalk{x_j,y} starting in time at least $h_2$, then the \twalk{x_1,y} obtained from following $W$ until reaching $(x_j,h_2)$, then proceeding with $Q$  is denoted by $x_1W(x_j,h_2)Qy$, or simply by  $x_1Wx_jQy$ when we are sure that $(x_j,h)$ occurs in $W$ and in $Q$ for some $h$.

Given a set of temporal paths $P_1,\ldots,P_q$, the \emph{union} of such paths is the temporal subgraph of $\mathcal{G}$ containing exactly the temporal edges of $P_1,\ldots,P_q$. Formally, if we denote such union by $\mathcal{G}'$, then $\mathcal{G}' = (G,\lambda')$ where $i\in \lambda'(e)$ if and only if $(e,i)\in E^T(P_j)$ for some $j\in [q]$.

Given a vertex $x$, we denote by $\delta^T_{\mathcal{G}}(x)$ the set of temporal edges incident in $x$, and by $d^T_{\mathcal{G}}(x)$ the value $\lvert \delta^T_{\mathcal{G}}(x)\rvert$. We omit ${\mathcal{G}}$ from the subscript if it its clear from the context.

Since some of the results depend on the application of our reduction from the strict to the non-strict scenario, we start by presenting such reduction.


\subsection{Temporal vertex-disjoint - Reduction from strict to non-strict}
\label{sec:strict_to_nonstrict}

In this section, we present a reduction from the strict model to the non-strict model that preserves \tvertex-disjoint paths and cuts. More formally, let $(H,\lambda)$ be a temporal (directed) graph.
We denote by $\mathcal{P}_{H,\lambda}(s,t)$ the set of non-strict \tpath{s,t}s in $(H,\lambda)$ and by $\mathcal{P}^*_{H,\lambda}(s,t)$ the set of non-strict \tpath{s,t}s in $(H,\lambda)$.

\begin{theorem}\label{thm:strict_to_nonstrict}
    Given a temporal (directed) graph $(G,\lambda)$ with lifetime $\tau$ and non-adjacent vertices $s,t$, there exists a temporal (directed) graph $(G',\lambda')$ {with lifetime $2\tau$} containing $s$ and $t$, and a bijective function $f:\mathcal{P}^*_{G,\lambda}(s,t)\rightarrow \mathcal{P}_{G',\lambda'}(s,t)$ such that:
    \begin{itemize}
        \item $\mathcal{P}\subseteq \mathcal{P}^*_{G,\lambda}(s,t)$ is a set of strict \tvertex-disjoint \tpath{s,t}s in $(G,\lambda)$ if and only if $f(\mathcal{P})$ is a set of non-strict \tvertex-disjoint \tpath{s,t}s in $(G',\lambda')$; and
        \item $S\subseteq (V(G)\setminus \{s,t\})\times [\tau]$ is a \tvcut{s,t} in $(G,\lambda)$ in the strict context  if and only if $f(S)$ is a \tvcut{s,t} in $(G',\lambda')$ in the non-strict context.
    \end{itemize}
\end{theorem}
\begin{proof}
    The transformation is an application of the so-called \emph{semaphore technique} used, for example,  in~\cite{CasteigtsCS22} and in~\cite{Costa00S23} to preserve reachability between different classes of temporal graphs.


    So, let $\mathcal{G}=(G,\lambda)$ be a temporal (directed) graph with lifetime $\tau$ and $s,t \in V(G)$ be non adjacent vertices. We construct the desired temporal (directed) graph $\mathcal{G}' = (G',\lambda')$ as follows.
    Start with $V(G')$ containing only $V(G)$. Then, for each temporal edge $(xy,i)$ in $\mathcal{G}$, we add two other vertices, denoted by $w^i_{xy}$ and $w^i_{yx}$  to $\mathcal{G}'$. Denote by $W$ the set of such vertices; we then have that $V(G')= V(G)\cup W$.
    Now, we consider the cases:
    \begin{itemize}
        \item $G$ is undirected: for each temporal edge $(xy,i)$ of $\mathcal{G}$, add to $\mathcal{G}'$ the temporal edges $(xw^i_{xy},2i)$, $(xw^i_{yx},2i+1)$, $(yw^i_{yx},2i)$, and $(yw^i_{xy},2i+1)$. Denote by $C^i_{xy}$ the cycle of size~4 formed by these edges.

        \item $G$ is directed: for each temporal edge $(xy,i)$ (note that here the order matters, i.e., $xy\neq yx$), add to $\mathcal{G}'$ the temporal edges $(xw^i_{xy},2i)$ and $(yw^i_{xy},2i+1)$. Denote by $C^i_{xy}$ the path on two vertices formed by these edges.
    \end{itemize}

    Observe that the arguments hold on both directed and undirected cases.
    First, we ensure the existence of the desired bijective function $f$. To see that $f$ is injective, just observe that two distinct \tpath{s,t}s in $\mathcal{G}$, $P$ and $Q$, differ in at least one edge; hence $f(P)\neq f(Q)$.
    So, let $P = (s,i_1,v_1,\ldots,i_k,v_k,i_{k+1},t)$ be a strict \tpath{s,t} in $(G,\lambda)$. We let $f(P)$ be the following \tpath{s,t} in $(G',\lambda')$:

    \[(s,2i_1, w^{i_1}_{sv_1}, 2i_1 +1, v_1,\ldots, 2i_k, w^{i_k}_{v_{k-1}v_k},2i_k +1,v_k,2i_{k+1},w^{i_{k}}_{v_kt}, 2{i_{k+1}}+1,t)\]

    Notice that $f(P)$ is an \tpath{s,t} in $(G',\lambda')$ if and only if $P$ is a strict \tpath{s,z} in $(G,\lambda)$. Therefore, $f$ is a well defined function. We now need to prove that $f$ is a bijection. To do this, observe that $(V(G),W)$ is a bipartition of $V(G')$. Therefore any walk in $G'$ alternates between such sets. In particular, any \tpath{s,t} in $(G',\lambda')$ has the following form:

    \[Q= (s,t_1,w_1,t_2,v_2,\dots,t_k,v_k,t_{k+1},w_{k+1},t_{k+2},t),\]

    where, for each $i \in [k+1]$, we have $w_i\in W$ and, for each $i\in \{2,\ldots,k\}$, we have that $v_i\in V(G)\setminus \{s,t\}$. Note that this infers a \tpath{s,t} in $\mathcal{G}$ because every $w\in W$ has degree two in $\mathcal{G}'$. It follows that $f$ is surjective and, hence, bijective as we wanted to prove.

    Now, we prove that the first item of the theorem holds. Suppose a strict temporal path $P$ has $(\alpha,x,\beta)$ as a subsequence\change{, with $x\in  V(G)$}. It means that $P$ \change{contains the temporal vertices $\{(x,\alpha+1),\ldots, (x,\beta)\}$}. Notice that, when transformed by $f$, we have the sequence $(2\alpha+1,x,2\beta)$ as a subsequence of $f(P)$  and  \change{$f(P)$ contains the temporal vertices $\{(x,2\alpha +1),\ldots,(x,2\beta)\}$ (recall that $f(P)$ is non-strict)}. Therefore, in order to show that disjoint \change{strict} \tpath{s,z}s are mapped into disjoint \change{non-strict} \tpath{s,z}s it is enough to show that:
    \[ \{\alpha_1+1,\dots,\beta_1\}\cap \{\alpha_2 +1,\dots, \beta_2\} = \emptyset \iff \{2\alpha_1+1,\dots,2\beta_1\}\cap \{2\alpha_2 +1,\dots, 2\beta_2\} = \emptyset.\]

    If we have $i$ in the first intersection, then $2i$ is in the second intersection. If we take $i$ in the second intersection, then $\lceil i/2 \rceil$ is in the first one.

    Finally, we prove the second item of the theorem's statement. For this, note that if a set of temporal vertices $S = \{(x_1, t_1),\dots, (x_k, t_k)\}$ intersects $V^T(P)$ for some strict \tpath{s,z} $P$ in $(G,\lambda)$, then $S' = \{(x_1,2t_1),\dots, (x_k,2t_k)\}$ intersects $V^T(f(P))$ in $(G',\lambda')$. On the other hand, if $S' = \{(x_1, t_1),\dots, (x_k, t_k)\}$ intersects $V^T(f(P))$ in $(G',\lambda')$ and $w^{\alpha}_{xy}\in S'$, then observe that we can replace $w^{\alpha}_{xy}$ by $x$ or $y$. Therefore, we can assume that $S' \subseteq V(G)\times [2\tau]$, in which case we can pick $S = \{(x_1,\lceil t_1/2 \rceil),\dots, (x_k,\lceil t_k/2 \rceil)\}$ as a cut in $(G,\lambda)$.
\end{proof}

Some of our results are a direct consequence from this reduction. We refer the reader to the $\star$ entries in \Cref{tab:summarywalks_nonstrict,tab:summarywalks_strict}.


\subsection{Menger's Theorem for \tvertex-disjoint walks}\label{sec:menger_walks}

As a part of our preliminary results, we now prove that Menger's Theorem always holds (and related parameters can be found in polynomial time), if walks instead of paths are considered, as stated below. Observe that this fills up column ``$=$'' and row ``\tvertex~- walk'' of \autoref{tab:summarywalks_nonstrict}. Additionally, as this is a positive result, by \autoref{thm:strict_to_nonstrict} we get that it also holds for strict temporal walks.

\begin{theorem}
    If $\mathcal{G}$ is a a temporal (directed) graph with lifetime $\tau$ and non-adjacent $s,t\in V(G)$, then $\tw(s,t)=\tc(s,t)$.
    Additionally, these values can be computed in polynomial time.
    \label{thm:tvertex_vwalks}
\end{theorem}
\begin{proof}
    We use what is known as the \emph{static expansion} of $\mathcal{G}$ (see e.g.~\cite{CLMS.21})\change{, which is also referred to as \emph{time-expanded graphs} in other contexts~\cite{Skutella2009}}. This is the directed graph having vertex set $V(G)\times [\tau]$, and edge set $E'\cup E''$, where $E'$ and $E''$ are defined below. When $G$ is an undirected graph, then we consider its directed version, i.e., where both $uv$ and $vu$ are arcs for every edge $uv$ of $G$.
    \[E' = \{(u,i)(u,i+1)\mid u\in V(G),i\in[\tau-1]\}.\]
    \[E'' = \{(u,i)(v,i)\mid uv\in E(G), i\in \lambda(uv)\}.\]

    Let $D$ denote the directed graph obtained from the static expansion of $\mathcal{G}$ by identifying all vertices $\{(s,i)\mid i\in [\tau]\}$ into a single vertex, $s$, and all vertices $\{(t,i)\mid i\in [\tau]\}$ into a single vertex, $t$. We prove that the maximum number of \tvertex-disjoint temporal $s,t$-walks in $\mathcal{G}$ is equal to the maximum number of vertex-disjoint $s,t$-paths in $D$, while the minimum size of a set $S\subseteq (V(G)\setminus \{s,t\})\times [\tau]$ that intersects every temporal $s,t$-walk in $\mathcal{G}$
    is equal to the minimum size of an $s,t$-cut in $D$, i.e., minimum $S\subseteq V(D)\setminus \{s,t\}$ intersecting all $s,t$-paths in $D$. The theorem thus follows by Menger's Theorem on static directed graphs, and the fact that computing these parameters in a static directed graph is largely known to be polynomial-time solvable (see e.g.~\cite{West.book}).

    First, given an $s,t$-path $P$ in $D$, we explain how to construct a temporal $s,t$-walk $P'$ in $\mathcal{G}$.
    So let $P = (\alpha_0 = s,e_1,\ldots,e_q,\alpha_q=t)$ be an $s,t$-path in $D$. We start with $P'$ being equal to $P$, and replace objects in $P'$ until we obtain a temporal $s,t$-walk in $\mathcal{G}$. So for each $i\in [q-1]$, write $\alpha_i$ as $(u_i,t_i)$.
    Also, for simplicity of notation, we consider $\alpha_0$ to be equal to $(s,t_1)$ and $\alpha_q$ to be equal to $(t,t_{q-1})$. Observe that $e_1 = s(u_1,t_1)$ and $e_q = (u_{q-1},t_{q-1})t$, i.e., $t_1$ and $t_{q-1}$ are the starting and finishing times of $P$, respectively.
    Now, for each $i\in [q]$, if $u_{i-1} = u_i$ (and hence $t_i = t_{i-1}+1$), remove $e_i$ and $\alpha_i$ from $P'$. And if $u_{i-1}\neq u_i$ (and hence $t_i = t_{i-1}$), then replace $e_i$ in $P'$ by $t_i$. Note that in the latter case, we get that $(u_{i-1}u_i,t_i)$ is a temporal edge of $\mathcal{G}$; call such fact (*).
    Observe that we are now left with a sequence that alternates temporal vertices and timesteps.
    Because of (*), and since $t_1\le t_2\le \ldots\le t_q$ as $(u,i)(v,j)$ is not an edge of $D$ whenever $i>j$, it suffices to replace each vertex $(u_i,t_i)$ in the sequence by simply $u_i$ in order to obtain a temporal walk in $\mathcal{G}$.
    One can verify that $V^T(P')\setminus(\{s,t\}\times[\tau]) = \{\alpha_1,\ldots,\alpha_{q-1}\} = V(P)\setminus \{s,t\}$. Additionally, observe that the backward transformation satisfying the same property can also be defined, i.e., given a temporal $s,t$-walk $P'$, we can construct an $s,t$-path $P$ in $D$ such that $V^T(P')\setminus (\{s,t\}\times[\tau]) = V(P)\setminus \{s,t\}$. This directly implies that the maximum number of \tvertex-disjoint temporal $s,t$-walks in $\mathcal{G}$ is equal to the maximum number of vertex-disjoint $s,t$-paths in $D$.

    So now suppose that $X\subseteq (V(G)\setminus \{s,t\})\times[\tau]$ is a minimum set intersecting every temporal $s,t$-walk in $\mathcal{G}$. Note that $X\subseteq V(D)$ and that, if there is an $s,t$-path not intersecting $X$ in $D$, then there is a temporal $s,t$-walk not intersecting $X$ in $\mathcal{G}$ by the previous paragraph, a contradiction.
    On the other hand, consider $X\subseteq V(D)\setminus\{s,t\}$ to be an $s,t$-separator in $D$. By construction, $X\subseteq (V(G)\setminus \{s,t\})\times [\tau]$. And, again by the previous paragraph, there cannot be a temporal $s,t$-walk in $\mathcal{G}$ not intersecting $X$.
\end{proof}


\subsection{Menger does not hold for \tvertex-disjoint paths}\label{sec:menger_does_not_hold}

We start by showing that the example ${\cal G}$ in \autoref{fig:tvertexpaths_counter} satisfies the properties claimed in the introduction, namely that $tp(s,t) = 2 < tpc(s,t) = 3$. The reader should observe that the same argument given in the proof below can be applied if, instead, ${\cal G}$ is the temporal directed graph obtained by replacing each edge $uv$ in the temporal graph of \autoref{fig:tvertexpaths_counter} by two arcs $\{(u,v),(v,u)\}$ with the same time labeling as $uv$.

\begin{proposition}
    Let ${\cal G} = (G,\lambda)$ be the temporal (directed) graph presented in \autoref{fig:tvertexpaths_counter}. Consider the non-strict model. Then $\tp(s,t) = 2 < \tpc(s,t) = 3$. 
    \label{prop:menger_counter}
\end{proposition}
\begin{proof}
    The reader should observe that the arguments also hold in the directed case.
    We first partition all \tpath{s,t}s into $\mathcal{P}^x$, $\mathcal{P}^y$, and $\mathcal{P}^{x,y}$, where $\mathcal{P}^I$ denotes the set of paths passing exactly by $I$ (and not by $\{x,y\}\setminus I$). Note that, by definition, this is indeed a partition of the paths.

    To get that $\tpc(s,t)=3$, observe that every path in $\mathcal{P} = \mathcal{P}^x\cup \mathcal{P}^y\cup \mathcal{P}^{xy}$ uses at least one of the temporal vertices $S = \{(x,1),(x,2),(y,1)\}$ (these are all the temporal neighbors of $s$). Hence, $\tpc(s,t)\le 3$. Additionally, note that $S\setminus \{(x,i)\}$, for each $i\in \{1,2\}$, is not a \tvcut{s,t} because of the path $(s,(sy,1),y,(yt,1),t)$. Finally, the path $(s,(sx,1),1,x, (xt,2),t)$ forces the existence of $(y,1)$ in the cut. It follows that $S$ is minimum.

    Now, since $\mathcal{P}^x\neq \emptyset$ and $\mathcal{P}^y\neq \emptyset$, we get $\tp(s,t)\ge 2$. To see that it is also at most 2, suppose by contradiction that $P_1,P_2,P_3$ are~3~\tvertex-disjoint \tpath{s,t}s. Because $s$ is incident to exactly 3 temporal edges, we can assume, without loss of generality that $(sx,1)\in E^T(P_1)$, $(sx,2)\in E^T(P_2)$ and $(sy,1)\in E^T(P_3)$. Now, because $(x,2)\in V^T(P_2)$, we get that $P_1$ must leave $x$ before timestep~2; hence $(xu,1)\in E^T(P_1)$. Also, since $(y,1)\in V^T(P_3)$, we get that $P_1$ cannot leave $u$ before timestep~2, and since $\lambda(uy) = \{1,2\}$, it must leave exactly in timestep~2. We get a contradiction since $P_3$ either uses $(yu,1)$ and thus intersects $P_1$ in $(u,1)$, or waits in $y$ until timestep 2 and intersects $P_1$ in $(y,2)$.
\end{proof}

Now, we generalize the above proposition to make the difference between the cut and the path parameters to be arbitrarily large.

\begin{proposition}
    For each fixed $k\geq 2$ and in the non-strict model, there exists a temporal (directed) graph ${\cal G} = (G,\lambda)$ and non-adjacent $s,t\in V(G)$ for which $\tpc(s,t)=\tp(s,t)+k$.
    \label{prop:menger_arbitrary_distance}
\end{proposition}
\begin{proof}
    To see that the difference between $\tp(s,t)$~and to $\tpc(s,t)$~can be arbitrarily large, just consider $k$ copies, ${\cal G}_1,\ldots,{\cal G}_k$ of the temporal graph given in \autoref{fig:tvertexpaths_counter}. For each $i\in [k]$, denote by $s_i,t_i$ the vertices corresponding to $s,t$ in ${\cal G}_i$, respectively. Then, let ${\cal G}^*$ be obtained by the disjoint union of ${\cal G}_1,\ldots,{\cal G}_k$ by identifying $s_1,\ldots,s_k$ into a single vertex, $s^*$, and also identifying $t_1,\ldots,t_k$ into a single vertex, $t^*$. By construction $tp_{{\cal G}^*}(s^*,t^*) = 2k$ while $tpc_{{\cal G}^*}(s^*,t^*)=3k$.
\end{proof}

Observe that the reduction in \autoref{thm:strict_to_nonstrict} can be applied to the constructions of \Cref{prop:menger_counter,prop:menger_arbitrary_distance} to obtain an analogous result in the strict model.

\begin{corollary}
    For each fixed $k\geq 2$ and in the strict model, there exists a temporal (directed) graph ${\cal G} = (G,\lambda)$ and non-adjacent $s,t\in V(G)$ for which $\tpc(s,t)=\tp(s,t)+k$.
    \label{cor:menger_arbitrary_distance_strict}
\end{corollary}


\section{Negative results}
\label{sec:negative}

In this section, we prove all the negative results in \autoref{tab:summarywalks_strict}. By \autoref{thm:strict_to_nonstrict}, the negative results in \autoref{tab:summarywalks_nonstrict} follow. We start with proving hardness for {\ktpaths}.

\begin{theorem}
    Consider the strict model. Given a temporal graph ${\cal G} = (G,\lambda)$, and non-adjacent $s,t\in V(G)$. Then \ktpaths\ is $\NP$-complete.
    \label{thm:paths_negative_undirected}
\end{theorem}
\begin{proof}
    To prove that the problem is in $\NP$, observe that checking whether a given set of paths $P_1,\cdots,P_k$ is indeed a set of \tvertex-disjoint \tpath{s,t}s can be done as follows. First, check whether each $P_i$ is indeed a temporal path.
    Then check whether each pair $P_i,P_j$ intersect in some temporal vertex.
    Observe that both checks can be done in polynomial-time, and hence we get that \ktpaths\ is $\NP$.

    To prove hardness, we reduce from $(2,2,3)$-\SAT, which consists of a variation of $3$-\SAT\ where each variable appears in exactly~4 clauses, twice positively and twice negatively. This is proved to be $\NP$-complete in~\cite{DA.19}. Given a formula $\phi$ on variables $x_1,\ldots,x_n$ and clauses $c_1,\ldots,c_m$, we first construct gadgets related to the variables. For each variable $x_i$, let $Q_i$ denote a cycle on 4 vertices, seen as a $2\times 2$ grid. For each $c,r\in [2]$, denote the vertex in column $c$ and row $r$ of $Q_j$ by $x^i_{r,c}$. The columns of $Q_i$ denote the positive appearances of $x_i$, while the rows denote the negative appearances of $x_i$.

    As a temporal graph can also be defined by a sequence of spanning subgraphs of a graph $G$, for simplicity, instead of describing function $\lambda$, we describe some gadget snapshots, then put them together in a sequence.
    For each clause $c_j$, we will have three snapshots related to $c_j$, called \emph{$i$-th clause snapshots}. Such snapshots will be alternated with what we call the \emph{$i$-th breaking snapshots}. In what follows, we consider that every snapshot is a spanning subgraph, so we refrain from saying that vertices are also being added to them, but describe simply the added edges. 

    We now describe the clause snapshots (observe \autoref{fig:clause_gadget}).
    For each $i\in [m]$, add two vertices $f_i$ and $\ell_i$ to $G$.
    For each variable $x_j$ appearing in $c_i$, we pick two vertices $v^1_{i,j},v^2_{i,j}$ of $Q_j$. If $x_j$ appears positively for the $h$-th time (recall that $h\in\{1,2\}$), let $v^1_{i,j},v^2_{i,j}$ be equal to $x^j_{1,h},x^j_{2,h}$, respectively (recall that the columns of $Q_j$ represent positive appearances). And if $x_j$ appears negatively for the $h$-th time, let $v^1_{i,j},v^2_{i,j}$ be equal to $x^j_{h,1},x^j_{h,2}$, respectively (since the rows of $Q_j$ represent negative appearances). Now, add the snapshots $G^1_i,G^2_i,G^3_i$ to $\mathcal{G}$ as represented in \autoref{fig:clause_gadget}. Finally, denote by $P_{i,j}$ the strict \tpath{f_i,\ell_i} passing by $v^1_{i,j},v^2_{i,j}$ and let the set $\{f_i\}\cup \bigcup_{x_j\in c_i}V(P_{i,j})$ be denoted by $S_i$.

    \begin{figure}[t]
        \centering
        \scalebox{0.8}{
            \begin{tikzpicture}[scale=1]
  \pgfsetlinewidth{1pt}
  \pgfdeclarelayer{bg}    
   \pgfsetlayers{bg,main} 
  
  \tikzset{vertex/.style={circle, minimum size=0.2cm, draw, inner sep=1pt, fill=black}}
  \tikzset{snapshot/.style ={draw=black!50, rounded corners, dashed, minimum height=8mm, minimum width=2.3cm} }
  \tikzset{subgraph/.style ={draw=black!50, circle, draw, minimum width=1cm, yscale=2, fill=white} }

  \node [label=-90:$G^1_i$, draw=black!50, rounded corners, dashed, minimum height=4cm, minimum width=4cm] at (0,0){};
  \node [label=-90:$G^2_i$, draw=black!50, rounded corners, dashed, minimum height=4cm, minimum width=4cm] at (5,0){};
  \node [label=-90:$G^3_i$, draw=black!50, rounded corners, dashed, minimum height=4cm, minimum width=4cm] at (10,0){};

  \node[vertex,label=180:$f_i$,fill=black!50] (f) at (0,1.7) {};  
  \node[vertex,label=-50:$x^2_{1,2}$,fill=black!50] (x21) at (0,0.7) {};  
  \node[vertex,label=120:$x^2_{2,2}$,fill=black!50] (x22) at (0,-0.7) {};  
  \node[vertex,label=150:$x^1_{1,1}$,fill=black!50] (x11) at (-1.1,0.7) {};  
  \node[vertex,label=150:$x^1_{2,1}$,fill=black!50] (x12) at (-1.1,-0.7) {};  
  \node[vertex,label=50:$x^3_{1,1}$,fill=black!50] (x31) at (1.1,0.7) {};  
  \node[vertex,label=50:$x^3_{1,2}$,fill=black!50] (x32) at (1.1,-0.7) {};  
  \node[vertex,label=180:$\ell_i$] (l) at (0,-1.7) {};  
    
  \begin{pgfonlayer}{bg}    
      \draw (f)--(x21); --(x22)--(l);
      \draw (f)--(x11);
      \draw (f)--(x31);
  \end{pgfonlayer}

  \begin{scope}[xshift=5cm]
    \node[vertex,label=180:$f_i$,fill=black!50] (f) at (0,1.7) {};  
    \node[vertex,label=-50:$x^2_{1,2}$,fill=black!50] (x21) at (0,0.7) {};  
    \node[vertex,label=120:$x^2_{2,2}$,fill=black!50] (x22) at (0,-0.7) {};  
    \node[vertex,label=150:$x^1_{1,1}$,fill=black!50] (x11) at (-1.1,0.7) {};  
    \node[vertex,label=150:$x^1_{2,1}$,fill=black!50] (x12) at (-1.1,-0.7) {};  
    \node[vertex,label=50:$x^3_{1,1}$,fill=black!50] (x31) at (1.1,0.7) {};  
    \node[vertex,label=50:$x^3_{1,2}$,fill=black!50] (x32) at (1.1,-0.7) {};  
    \node[vertex,label=180:$\ell_i$] (l) at (0,-1.7) {};  
    \begin{pgfonlayer}{bg}    
      \draw (x21)--(x22);
      \draw (x11)--(x12);
      \draw (x31)--(x32); 
    \end{pgfonlayer}
  \end{scope}

  \begin{scope}[xshift=10cm]
    \node[vertex,label=180:$f_i$,fill=black!50] (f) at (0,1.7) {};  
    \node[vertex,label=-50:$x^2_{1,2}$,fill=black!50] (x21) at (0,0.7) {};  
    \node[vertex,label=120:$x^2_{2,2}$,fill=black!50] (x22) at (0,-0.7) {};  
    \node[vertex,label=150:$x^1_{1,1}$,fill=black!50] (x11) at (-1.1,0.7) {};  
    \node[vertex,label=150:$x^1_{2,1}$,fill=black!50] (x12) at (-1.1,-0.7) {};  
    \node[vertex,label=50:$x^3_{1,1}$,fill=black!50] (x31) at (1.1,0.7) {};  
    \node[vertex,label=50:$x^3_{1,2}$,fill=black!50] (x32) at (1.1,-0.7) {};  
    \node[vertex,label=180:$\ell_i$] (l) at (0,-1.7) {};  
    \begin{pgfonlayer}{bg}    
     \draw (x22)--(l);
     \draw (x12)--(l);
     \draw (x32)--(l);
    \end{pgfonlayer}
  \end{scope}
    
    


  \end{tikzpicture}
        }
        \caption{Example of clause snapshots. We take $c_i = x_1\vee x_2\vee \neg x_3$, where $x_1$ occurs for the 1st time, $\neg x_2$ for the 2nd time, and $\neg x_3$ for the 1st time. Grey vertices denote $S_i$.}
        \label{fig:clause_gadget}
    \end{figure}

    Now, we construct the breaking snapshots. They will appear after each set of clause snapshots. So consider $i\in [m]$, and denote by $s_i$ the number of literals in $c_i$. The first two breaking snapshots that appear after $G^3_i$ are denoted by $B^1_i,B^2_i$ and are formed by $2s_i+1$ strict $s,t$-paths, each passing by a distinct vertex of $S_i$ (see \autoref{fig:breaking_gadget}).
    Finally, we add the third breaking snapshot. It appears after $B^2_i$, is denoted by $L_i$, and is formed by the edge $\ell_if_{i+1}$.

    \begin{figure}[t]
        \centering
        \scalebox{0.8}{
            \begin{tikzpicture}[scale=1]
  \pgfsetlinewidth{1pt}
  \pgfdeclarelayer{bg}    
   \pgfsetlayers{bg,main} 
  
  \tikzset{vertex/.style={circle, minimum size=0.2cm, draw, inner sep=1pt, fill=black}}
  \tikzset{snapshot/.style ={draw=black!50, rounded corners, dashed, minimum height=8mm, minimum width=2.3cm} }
  \tikzset{subgraph/.style ={draw=black!50, circle, draw, minimum width=1cm, yscale=2, fill=white} }

  \node [label=-90:$B^1_i$, draw=black!50, rounded corners, dashed, minimum height=4cm, minimum width=4cm] at (0,0){};
  \node [label=-90:$B^2_i$, draw=black!50, rounded corners, dashed, minimum height=4cm, minimum width=4cm] at (5,0){};
  \node [label=-90:$L_i$, draw=black!50, rounded corners, dashed, minimum height=4cm, minimum width=4cm] at (10,0){};
      
  \begin{pgfonlayer}{bg}    
      
  \end{pgfonlayer}

  \node[vertex,label=180:$s$] (s) at (0,1.7) {};  
  \node[label=20:$S_i$, draw=black!50, rounded corners, dashed, minimum height=1cm, minimum width=3.5cm] at (0,0){};
  \node[vertex,fill=black!50] (f) at (-1.4,0) {};
  \node[vertex,fill=black!50] (x11) at (-0.6,0) {};
  \node[vertex,fill=black!50] (x32) at (1.4,0) {};
  \node at (0.5,0) {$\ldots$};
  \node[vertex,label=180:$t$] (t) at (0,-1.7) {};  
    
  \begin{pgfonlayer}{bg}    
      \draw (s)--(f) (s)--(x11) (s)--(x32);
  \end{pgfonlayer}

  \begin{scope}[xshift=5cm]
    \node[vertex,label=180:$s$] (s) at (0,1.7) {};  
    \node[label=20:$S_i$, draw=black!50, rounded corners, dashed, minimum height=1cm, minimum width=3.5cm] at (0,0){};
    \node[vertex,fill=black!50] (f) at (-1.4,0) {};
    \node[vertex,fill=black!50] (x11) at (-0.6,0) {};
    \node[vertex,fill=black!50] (x32) at (1.4,0) {};
    \node at (0.5,0) {$\ldots$};
    \node[vertex,label=180:$t$] (t) at (0,-1.7) {};  
    
    \begin{pgfonlayer}{bg}    
      \draw (f)--(t) (x11)--(t) (x32)--(t);
    \end{pgfonlayer}

  \end{scope}

  \begin{scope}[xshift=10cm]
   \node[vertex,label=180:$\ell_i$] (l) at (0,1) {};
   \node[vertex,label=180:$f_{i+1}$] (f) at (0,-1) {};
   \begin{pgfonlayer}{bg}    
       \draw (l)--(f);
   \end{pgfonlayer}  
  \end{scope}
  
  \end{tikzpicture}
        }
        \caption{Breaking snapshots. Grey vertices denote $S_i$.}
        \label{fig:breaking_gadget}
    \end{figure}

    To conclude the construction, we add a 0-th snapshot containing simply the $sf_1$ and and a final snapshot containing the edge $f_{m+1}t$; denote these $G_0$ and $G_f$. Let ${\cal G}$ be the temporal graph formed by the sequence $(G^1_1,G^2_1,G^3_1,B^1_1,B^2_1,L_1,G^1_2,\ldots,L_m,G_f)$.
    Note that each pair of breaking snapshots related to a certain clause $c_i$ gives us $2s_i+1$ \tvertex-disjoint \tpath{s,t}s, those containing as internal \tvertex only one element of $S_i$. This already gives us $N = 2\sum_{i=1}^{m}s_i + m$ such disjoint paths. Now, we need to prove that $\phi$ is satisfiable if and only if ${\cal G}$ has $N+1$ \tvertex-disjoint \tpath{s,t}s. The idea is that there will be always at least $N$ \tvertex-disjoint strict \tpath{s,t}s and that the extra one will exist if and only if $\phi$ has a satisfying assignment.

    First, suppose that $\phi$ has a satisfying assignment. We first construct $N$ \tvertex-disjoint \tpath{s,t}s whose temporal edges are contained within the breaking snapshots. For simplicity, given a snapshot $S$, we denote by $S$ itself the time step in which snapshot $S$ is used. We then start with the following clearly \tvertex-disjoint paths.
    For each $i\in [m]$ and each $u\in S_i$, we pick $P_u = (s,B^1_i,u,B^2_i,t)$.This gives us exactly $N$ such paths, as desired.
    Now, we construct the extra strict \tpath{s,t}. For this, we cannot intersect any of the breaking snapshots. We then construct an $s,t$-walk $P$ containing some of the $f_i,\ell_i$-paths in $G_i$, for every $i\in [m]$, then we prove that $P$ is the desired temporal path. Intuitively, the picked path is formed by the concatenation of subpaths contained within the clause snapshots. So, let $P$ start with $(s,G_0,f_1)$. At the $i$-th iteration, our partial path will always have as endpoint $f_i$. Now, suppose we are at the beginning $i$-th iteration, $i\in [m]$, and let $x_j$ be a variable that validates $c_i$.
    Add to $P$ the \tpath{f_i,\ell_i} path, $P_{i,j}$, contained in $G^1_i,G^2_i,G^3_i$ and the path $(\ell_i,L_i,f_{i+1})$. Finally, at the end of the $m$-th iteration, $P$ has $f_{m+1}$. We then simply add the path $(f_{m+1},G_f,t)$.

    By construction, one can see that $P$ is a temporal $s,t$-walk that does not intersect any of the previously picked \tpath{s,t}s. So, it remains to argue that $P$ is actually a temporal path, i.e., that $P$ does not contain two occurrences of the same vertex of $G$.
    Suppose, without loss of generality, that $x^1_{1,1}$ is a vertex that appears twice in $P$ (note that this can be done since no particular order of the variables is assumed).
    Note that the two positive occurrences are denoted in the columns and the two negative occurrences in the rows. This means that $x^1_{1,1}$ is added during the iteration of a clause which it appears positively, then again during the iteration where it appear negatively. This is a contradiction as we always pick literals that validate the clauses and because $x_1$ must be either true or false.

    Now, suppose that $P_1,\ldots,P_{N+1}$ are \tvertex-disjoint \tpath{s,t}s. Because $d^T(s) = d^T(t) = N+1$, we know that the $N$ $s,t$-paths contained in the breaking snapshots must be part of these disjoint paths.
    So, let $P = P_1$ the path passing by $sf_1$ and $f_{m+1}t$ (in the first and last snapshot, respectively).
    Observe that for each $j\in [n]$, if $P$ contains some column of $Q_j$, then $P$ cannot contain any row of $Q_j$, and vice-versa, as every pair of row and column intersect and this would imply a repetition of some vertex \change{in $P$ (recall that $P$ is a \emph{path}, not a walk)}. Therefore, we set $x_j$ to true if $P$ contains some column of $Q_j$, and to false otherwise. This is a satisfying assignment for $\phi$ since for every $i\in [m]$,  $P$ must contain some of the $f_i,\ell_i$-paths in $G_i$ as otherwise $P$ would intersect some $s,t$-path in $B_i$.
\end{proof}

We now use the same construction to prove hardness of the cut problems.

\begin{theorem}
    Consider the strict model. Given a temporal (directed) graph ${\cal G} = (G,\lambda)$ with lifetime $\tau$, non-adjacent vertices $s,t\in V(G)$, and $S\subseteq (V(G)\setminus \{s,t\})\times [\tau]$,  then
    \begin{enumerate}
        \item Deciding whether $S$ is a \tvcut{s,t} in ${\cal G}$ is co-$\NP$-complete; and
        \item \htcuts\ is co-$\NP$-hard.
    \end{enumerate}
    \label{thm:decide_if_cut}
\end{theorem}
\begin{proof}
    Consider  the same construction as before. Note that, for each $i\in [m]$ and $u\in S_i$, the path passing only by snapshots $B^1_i,B^2_i$, containing as unique internal vertex $(u,B^2_i)$, enforces us to include it in any \tvcut{s,t}.
    So, let $\mathcal{S} = \bigcup_{i=1}^m \{(u,B^2_i)\mid u\in S_i\}$.
    We argue that $\phi$ is satisfiable if and only if $S$ is not a \tvcut{s,t}. This means that deciding whether $S$ is a \tvcut{s,t} is co-\NP-hard. Since an \tpath{s,t} not intersecting $S$ is a certificate for $S$ not being a cut that can be checked in polynomial time, we
    get that Item~\lipItem{1.} of our theorem follows, i.e., the problem is co-\NP-complete.
    To prove hardness, just notice $S$ is not a \tvcut{s,t} if and only if there exists an \tpath{s,t} using $(sf_1,1)$ and passing through all the clauses, in which case ${\cal G}$ has $N+1$ \tvertex-disjoint \tpath{s,t}s. \change{Observe that} Item~\lipItem{1.} of \autoref{thm:decide_if_cut} then follows from the previous proof.

    Finally, to see the hardness of {\htcuts}, just observe that if any $S'\subseteq (V(G)\setminus \{s,t\})\times [\tau]$ is such that $S\nsubseteq S'$, then $S'$ cannot be a \tvcut{s,t} as every $(u,i)\in S$ is adjacent to both $s$ and $t$ in timestep~$i$, i.e., $(s,i,u,i,t)$ is an  \tpath{s,t} not intersecting $S'$. Therefore, ${\cal G}$ has a \tvcut{s,t} of size $N$ if and only if $S$ is a \tvcut{s,t} in ${\cal G}$.
\end{proof}

Finally, we prove that {\ktpaths} is $\NP$-complete on directed temporal graphs even if $k\ge 3$ and $\tau\ge 3$ are fixed. This means that it is para-$\NP$-complete when parameterized either by the size of the solution or by the lifetime of ${\cal G}$. The proof cannot be adapted to strict paths since spreading the snapshots to make the paths strict may make them overlapping.

\begin{theorem}\label{thm:paths_negative_directed}
    Consider the non-strict model. Given a temporal directed graph $\mathcal{G} = (G,\lambda)$, and non-adjacent $s,t\in V(G)$. Then, \ktpaths\ is $\NP$-complete for fixed values of $k\ge 3$ when $G$ is a directed graph, even if $\tau=3$.
\end{theorem}
\begin{proof}
    \change{Given a temporal (directed) graph ${\cal G} = (G,\lambda)$, and non-adjacent $s,t\in V(G)$, we want to prove that \ktpaths\ is $\NP$-complete for fixed values of $k\ge 3$ when $G$ is a directed graph, even if $\tau=3$. }
    We already know that the problem is in $\NP$. To prove hardness, we make a reduction from $2$-\textsc{Linkage} to prove \NP-completeness for $k=3$. For bigger values of $k$, it suffices to add to the constructed instance $k-3$ distinct new vertices, all simultaneously adjacent to $s$ and $t$ in timestep~2. Problem $2$-\textsc{Linkage} consists of, given a directed graph $D = (V,A)$ and two pairs of non-adjacent vertices $x_1,y_1$ and $x_2, y_2$ of $V$, deciding whether there are vertex-disjoint paths from $x_1$ to $y_1$ and from $x_2$ to $y_2$ in $D$. This problem is known to be $\NP$-complete~\cite{FORTUNE1980111}. We remark that we can always assume that $x_1$ and $x_2$ are sources and that $y_1$ and $y_2$ are sinks.
    So, let $(D, x_1, y_1, x_2, y_2)$ be an instance of $2$-\textsc{Linkage}, where $D = (V,A)$. We construct an equivalent instance of $\geq 3$ $\textsc{\tvertex-disjoint Paths}$ on a temporal directed graph $\mathcal{G} = (G, \lambda)$ as follows (see \autoref{fig:reduction-2-linkage} to follow the construction).
    Let $X = V(D) \setminus \{x_1, y_1, x_2, y_2\}$.
    First, we add to $G$ every vertex of $X$, a vertex $w_{1,2}$ associated with $x_1$ and $y_2$, a vertex $w_{2,1}$ associated with $x_2$ and $y_1$, and new vertices $s$ and $t$. Also, add to $G$ arcs $\{(s,w_{1,2}),(s, w_{2,1}),(w_{1,2}, w_{2,1}),(w_{1,2},t),(w_{2,1}, t)\}$ and let: $\lambda(s,w_{1,2}) = \{1,2\}$, $\lambda(s,w_{2,1}) = \{1\}$, $\lambda(w_{1,2},w_{2,1}) = \{2\}$, $\lambda(w_{1,2},t) = \{3\}$, and $\lambda(w_{2,1},t) = \{2,3\}$.
    Finally, do as follows:
    \begin{enumerate}[(i)]
        \item for each $e \in A(D)$ with both endpoints in $X$, add to $G$ a copy of $e$ with $\lambda(e) = \{1\}$;
        \item for each $(x_1, v) \in A(D)$ add an arc $e = (w_{1,2}, v)$ with $\lambda(e) = \{1\}$;
        \item for each $(x_2, v) \in A(D)$ add an arc $e = (w_{2,1}, v)$  with $\lambda(e) = \{1\}$;
        \item for each $(v, y_1) \in A(D)$ add an arc $e = (v, w_{2,1})$  with $\lambda(e) = \{3\}$; and
        \item for each $(v, y_2) \in A(D)$ add an arc $e = (v, w_{1,2})$  with $\lambda(e) = \{3\}$.
    \end{enumerate}

    \begin{figure}[htb]
        \centering
        \scalebox{0.7}{
            \begin{tikzpicture}[scale=1]
                \pgfsetlinewidth{1pt}
                \pgfdeclarelayer{bg}
                \pgfsetlayers{bg,main}

                \tikzset{vertex/.style={circle, minimum size=0.2cm, draw, inner sep=1pt, fill=black}}
                \tikzset{snapshot/.style ={draw=black!50, rounded corners, dashed, minimum height=8mm, minimum width=2.3cm} }
                \tikzset{subgraph/.style ={draw=black!50, circle, draw, minimum width=1cm, yscale=2, fill=white} }

                \node [label=-90:$G_1$, draw=black!50, rounded corners, dashed, minimum height=7cm, minimum width=5cm] at (-0.5,0){};
                \node [draw=black!50, minimum height=3cm, minimum width=2cm] at (0,0){$G[X]$};
                \node [draw=black!50, dotted] at (0,1){$N^+(x_1)$};
                \node [draw=black!50, dotted] at (0,-1){$N^+(x_2)$};

                \node[vertex,label=0:$w_{1,2}$] (w12) at (0,2.5) {};
                \node[vertex,label=0:$w_{2,1}$] (w21) at (0,-2.5) {};
                \node[vertex,label=180:$s$] (s) at (-2,0) {};

                \begin{pgfonlayer}{bg}    
                    \path[->,>=latex] (w12) edge (0,1.3) (w12) edge (-0.5,1.3) (w12) edge (0.5,1.3);
                    \path[->,>=latex] (w21) edge (0,-1.3) (w21) edge (-0.5,-1.3) (w21) edge (0.5,-1.3);
                    \path[->,>=latex] (s) edge [out=90,in=180] (w12) (s) edge [out=-90,in=180] (w21);
                \end{pgfonlayer}

                \node [label=-90:$G_2$, draw=black!50, rounded corners, dashed, minimum height=7cm, minimum width=5cm,xshift=6cm] at (-0.5,0){};
                \node[vertex,label=180:$s$,xshift=6cm] (s) at (-2,0) {};
                \node[vertex,label=0:$t$,xshift=6cm] (t) at (1,0) {};
                \node[vertex,label=90:$w_{1,2}$,xshift=6cm] (w12) at (-0.5,1.5) {};
                \node[vertex,label=-90:$w_{2,1}$,xshift=6cm] (w21) at (-0.5,-1.5) {};
                \begin{pgfonlayer}{bg}    
                    \path[->,>=latex] (s) edge (w12) (w12) edge (w21) (w21) edge (t);
                \end{pgfonlayer}

                \node [label=-90:$G_3$, draw=black!50, rounded corners, dashed, minimum height=7cm, minimum width=5cm,xshift=12cm] at (-0.5,0){};
                \node[vertex,label=0:$t$,xshift=12cm] (t) at (1,0) {};
                \node [draw=black!50, minimum height=3cm, minimum width=2cm,xshift=12cm] at (-1,0){$X$};
                \node [draw=black!50, dotted,xshift=12cm] at (-1,1){$N^-(y_2)$};
                \node [draw=black!50, dotted,xshift=12cm] at (-1,-1){$N^-(y_1)$};
                \node[vertex,label=180:$w_{1,2}$,xshift=12cm] (w12) at (-1,2.5) {};
                \node[vertex,label=180:$w_{2,1}$,xshift=12cm] (w21) at (-1,-2.5) {};
                \begin{pgfonlayer}{bg}    
                    \path[->,>=latex,xshift=12cm] (-1,1.3) edge (w12) (-1.5,1.3) edge  (w12) (-0.5,1.3) edge (w12);
                    \path[->,>=latex,xshift=12cm] (-1,-1.3) edge (w21) (-1.5,-1.3) edge  (w21) (-0.5,-1.3) edge (w21);
                    \path[->,>=latex] (w12) edge [out=0,in=90] (t) (w21) edge [out=0,in=-90] (t);
                \end{pgfonlayer}

            \end{tikzpicture}

        }%
        \caption{Construction used in the proof of Item~\lipItem{2.} of \autoref{thm:paths_negative_directed}. Notice that all edges of $G[X]$ appear only in $G_1$ and that $X$ in $G_3$ is an independent set. $N^+(u),N^-(u)$ denote, respectively, the out-neighborhood and the in-neighborhood of a vertex $u$.}
        \label{fig:reduction-2-linkage}
    \end{figure}

    To conclude, we prove that $(D,x_1,y_1,x_2,y_2)$ is a positive instance of $2$-\textsc{Linkage} if and only if there are at least 3 \tvertex-disjoint \tpath{s,t}s in ${\cal G}$.
    Because in this proof we work with paths in static graphs, as well as temporal paths, we use the complete designation of the latter. The idea of the proof is that we will force the $x_1,y_1$-path and the $x_2,y_2$-path in $G$ to wait on the neighbors of $y_1$ and $y_2$, respectively, in order to make way for a third temporal $s,t$-path in $\mathcal{G}$.

    Observe first that we can suppose that $(x_i,y_i) \notin A(D)$, for each $i\in [2]$, as otherwise, if say $(x_1,y_1)\in A(D)$, then we can solve $2$-\textsc{Linkage} in polynomial-time by removing arc $(x_1,y_1)$ and vertices $\{x_1,y_1\}\setminus \{x_2,y_2\}$ from $D$, and testing whether the obtained graph contains an $x_2,y_2$-path. Such path will surely be disjoint from the $x_1,y_1$-path formed just by that arc.

    Now, for each $i\in [2]$, let $P_i$ be an $x_i,y_i$-path in $D$, and suppose that $P_1$ and $P_2$ are internally disjoint. Write $P_1$ as $(x_1=u_0,u_1,\cdots,u_q=y_1)$ and let $P'_1$ be the temporal $w_{1,2},w_{2,1}$-walk obtained by following $P_1$ within $G_1$ until $u_{q-1}$, then using $(u_{q-1}w_{2,1},3)$. Formally,  $P'_1 = (w_{1,2},1,u_1,\ldots,1,u_{q-1},3,w_{2,1})$. We argue that $P'_1$ is a temporal $w_{1,2},w_{2,1}$-path in ${\cal G}$. Indeed, since $P_1$ is a path and all temporal vertices of $P'_1$ are contained in timestep~1 except $w_{2,1}$, we get that the only way it could repeat a vertex is if $(w_{2,1},1)\in V^T(P'_1)$, which can only happen if $x_2\in V(P_1)$, contradicting the disjointness of $P_1$ and $P_2$. Clearly $P'_1$ can be extended to a temporal $s,t$-path, $Q_1$, by adding $(sw_{1,2},1)$ to the beginning and $(w_{2,1}t,3)$ to the end. A similar argument can be applied to obtain a second temporal $s,t$-path, $Q_2$, and one can see that $Q_1$ and $Q_2$ are \tvertex-disjoint as $P_1$ and $P_2$ are internally vertex-disjoint. Finally, let $Q_3 = (s,2,w_{1,2},2,w_{2,1},2,t)$ be our third temporal $s,t$-path.
    Observe that, because $(x_i,y_i)\notin A(D)$ for each $i\in [2]$, we get that neither $Q_1$ nor $Q_2$ uses $w_{1,2}$ nor $w_{2,1}$ at time $2$, thus giving us that $Q_3$ is \tvertex-disjoint from both $Q_1$ and $Q_2$.

    Assume now that there are \tvertex-disjoint temporal $s,t$-paths $Q_1, Q_2$, and $Q_3$ in $\mathcal{G}$.
    Since there are exactly 3~temporal arcs entering $t$, we can suppose, without loss of generality, that $Q_1$ uses $(w_{1,2}t,3)$, $Q_2$ uses $(w_{2,1}t,3)$ and $Q_3$ uses $(w_{2,1}t,2)$. Similarly, we get that each of the 3 arcs leaving $s$ must be contained in a distinct such path. Observe that this means also that each copy of $w_{1,2}$ and each copy of $w_{2,1}$ is used by exactly one of these 3 paths, and therefore none of them can wait in neither $w_{1,2}$ nor in $w_{2,1}$. As a consequence we get that $Q_3$ must be equal to $(s,2,w_{1,2},2,w_{2,1},2,t)$.
    Now, since $Q_1$ and $Q_2$ are \tvertex-disjoint they do not intersect in $X$ at time $1$ and we can follow those two paths to construct vertex-disjoint $x_i,y_i$-paths in $D$, for each $i \in [2]$.
    By replacing the temporal arc leaving $w_{1,2}$ at time $1$ by the associated arc of $D$ leaving $x_1$, following $Q_1$ inside of $X$, and replacing the temporal arc reaching $w_{2,1}$ at time $3$ by the associated arc of $D$ reaching $y_1$, we have constructed an $x_1, y_1$-path in $D$.
    We can construct an $x_2, y_2$-path in $D$ with similar steps, and the results follows by the disjointness of $Q_1$ and $Q_2$.
\end{proof}

\section{Positive Results}\label{sec:positive_results}

In this section, we present all of our positive results. The first, most basic one, is the following.

\begin{proposition}\label{prop:kh_1}
    Given a temporal (directed) graph, solving {\textsc{(Directed)}\ktpaths} for $k=1$ and solving {\textsc{(Directed)} \htcuts} for $h=1$ can be done in polynomial time.
    \label{prop:kequals1_easy}
\end{proposition}
\begin{proof}
    Because a walk always contains a path, one can simply apply any temporal connectivity algorithm (e.g.~\cite{Wu.etal.16} solves it in time $O(m\tau)$) to decide whether there is a temporal $s,t$-walk. This holds for all variations, strict and non-strict, directed or undirected. If so, then there exists at least one path and the answer to {\textsc{(Directed)}\ktpaths} with $k=1$ and for {\textsc{(Directed)} \htcuts} with $h=1$ are both yes. Otherwise, the answer to both problems is no.
\end{proof}

Now, observe that the problems defined in the introduction does not tell us anything about equality of parameters. In fact, we have seen in \autoref{sec:menger_does_not_hold} that these parameters can be arbitrarily far apart. Our next section tells us that if the maximum number of paths is~1, then it must be also the case for the minimum size of a cut.

\subsection{Menger's Theorem for \tvertex-disjoint paths when $k=1$}\label{sec:mengerpaths}

In this section we prove our main result, namely that a version of Menger holds for \tvertex-disjoint paths and cuts, but only if the maximum number of paths is equal to~1. Our proof considers the non-strict case. This is enough to get that it holds also for the strict case based on \autoref{thm:strict_to_nonstrict}.

Observe that the result can be rewritten as: $tp(s,t)=1$ if and only if $tpc(s,t)=1$. To see that $tpc(s,t)=1$ implies $tp(s,t)=1$  just observe that $tpc(s,t)\ge tp(s,t)$ and that if no \tpath{s,t} exists, then we get $tpc(s,t)=0$. It thus remains to prove that if $tp(s,t)=1$, then $tpc(s,t)=1$. \change{We do this by contraposition, i.e., we prove that if $tpc(s,t)>1$, then $tp(s,t)>1$. In turn, we prove this by contradiction, that is, we suppose the that $tpc(s,t)>1$ and $tp(s,t)=1$. By investigating properties of a minimal temporal graph with such property, we arrive to a contradiction.}

\begin{theorem}\label{thm:menger}
    Consider the non-strict case. Let $\mathcal{G} = (G,\lambda)$ be a temporal (directed) graph and consider non-adjacent $s,t\in V(G)$. Then, $\tp(s,t)=1$  iff $\tpc(s,t)=1$.
\end{theorem}

We start by giving a general idea of a proof of the classical Menger's Theorem, since our proof has some resemblance with it. This proof of Menger's Theorem is done by induction on the number of vertices. Consider then a graph $G$, non-adjacent vertices $s,t$ and a minimum $s,t$-cut $S$. Let $H_s$ and $H_t$ be the components of $G-S$ containing $s$ and $t$, respectively. Then, two graphs are constructed: $G_1$ obtained from $G[V(H_s)\cup S]$ by adding a new vertex $t'$ adjacent to all vertices of $S$; and $G_2$ obtained from $G[V(H_t)\cup S]$ by adding a new vertex $s'$ adjacent to all vertices of $S$. If $S$ is not contained in $N(s)$ or in $N(t)$, then $G_1$ and $G_2$ are smaller than $G$. We can then apply induction to obtain $k = \lvert S\rvert$ vertex-disjoint $s,t'$-paths in $G_1$ and $k$ vertex-disjoint $s',t$-paths in $G_2$. This gives us disjoint paths from $s$  to $S$ that can be combined with paths from $S$ to $t$ to form $k$ disjoint paths in $G$. Finally, if either $G_1$ or $G_2$ is not smaller than $G$, then other techniques are used to obtain the paths. In our proof, given a \tvcut{s,t} $S$, we try to obtain \tvertex-disjoint paths from $s$ to $S$ and from $S$ to $t$ and then combine them. However, two main difficulties arise. First, we cannot always combine an \tpath{s,u} $P$ with a \tpath{u,t} $P'$, as $P$ might arrive in $u$ too late in order to move on with $P'$. Second, even if $P$ and $P'$ can be combined, it might not lead to a path since some vertex $v$ might be an internal vertex of $P$ and of $P'$. Observe that, in this case, the $s,t$-path obtained by waiting on $v$ until its occurrence in $P'$ might produce a path that intersects the other ones. We now move on to presenting our proof. We use a minimum counter-example instead of induction, as we believe the proof becomes clearer.

We say that ${\cal G} = (G,\lambda), s,t$ is a \emph{minimal counter-example} (for \autoref{thm:menger}) if $st\notin E(G)$, $tp_{{\cal G}}(s,t)=1<tpc_{{\cal G}}(s,t)$, and
${\cal G}$ does not contain a smaller counter-example, i.e.: for every ${\cal H} = (H,\lambda')\subset (G,\lambda)$, if $s',t'\in V(H)$ are such that $s't'\notin E(H)$ and $tp_{{\cal H}}(s',t')=1$, then $tpc_{{\cal H}}(s',t')=1$. It is \emph{minimum} if, among all minimal counter-examples, it minimizes the value $\lvert V(G)\rvert+\lvert E^T({\cal G})\rvert$.
Observe that, by definition of minimal counter-example, it follows that $tpc_{\cal G}(s,t)\ge 2$. In what follows, we omit the subscript whenever ${\cal G}$ is clear from the context.
We start by proving the following proposition, which will be very useful throughout the proof.

\begin{proposition}\label{prop:remove_tempedge}
    Let $\mathcal{G},s,t$ be a minimum counter-example. For every temporal edge $(xy,i)\in E^T({\cal G})$, we have $tpc_{{\cal G}-(xy,i)}(s,t) = tp_{\mathcal{G}-(xy,i)}(s,t)=1$.
\end{proposition}
\begin{proof}
    As $st\notin E(\mathcal{G})$, we can suppose without loss of generality that $x\notin \{s,t\}$. Note that $tp_{{\cal G}-(xy,i)}(s,t)=1$ as otherwise we would get that $\{(x,i)\}$ is a \tvcut{s,t} in $\mathcal{G}$.
    And since ${\cal G}-(xy,i)$ is not a counter-example, as it is smaller than $\mathcal{G}$, it follows that $tpc_{{\cal G}-(xy,i)}(s,t) = tp_{\mathcal{G}-(xy,i)}(s,t)=1$.
\end{proof}

We now prove that a minimum \tvcut{s,t} in ${\cal G}$ must always contain exactly two temporal vertices.

\begin{lemma}\label{lem:tpc=2}
    Let $\mathcal{G},s,t$ be a minimal counter-example. Then $tpc(s,t) = 2$.
\end{lemma}
\begin{proof}
    Denote the lifetime of $\mathcal{G}$ by $\tau$. Let $(uv,i)$ be any temporal edge in ${\cal G}$; it exists as otherwise $tp_{\mathcal{G}}(s,t) = 0 = tpc_{\mathcal{G}}(s,t)$. Let $\mathcal{G}' = \mathcal{G}-(uv,i)$. Since $st\notin E(\mathcal{G})$, we can suppose, without loss of generality, that $v\notin \{s,t\}$. By \autoref{prop:remove_tempedge}, let
    $\{(x,i_x)\}\subseteq V(\mathcal{G}')\times  [\tau]$ be a minimum \tvcut{s,t} in $\mathcal{G}'$. Since any \tpath{s,t} in $\mathcal{G}$ not contained in $\mathcal{G}'$ must pass by $(uv,i)$, we get that $\{(x,i_x),(v,i)\}$ is a \tvcut{s,t} in $\mathcal{G}$. Therefore we get $tpc(s,t)\le 2$. The lemma follows as we also have $1 = tp(s,t) < tpc(s,t)$.
\end{proof}

Our first important lemma is the following, which shows that the two temporal vertices in a \tvcut{s,t} cannot be copies of the same vertex of $\mathcal{G}$.

\begin{lemma}\label{lem:nocutvertex}
    Let $\mathcal{G},s,t$ be a minimum counter-example. Then there is no $u\in V(\mathcal{G})\setminus \{s,t\}$ such that every \tpath{s,t} contains $u$.
\end{lemma}
\begin{proof}

    Recall that $pc(s,t)$ denotes the minimum size of a set $S\subseteq V(\mathcal{G})\setminus\{s,t\}$ that intersects every \tpath{s,t}. We prove that $pc(s,t)\ge tpc(s,t)$ and the lemma follows from \autoref{lem:tpc=2}. By contradiction, suppose that $u\in V(\mathcal{G})\setminus\{s,t\}$ is such that every \tpath{s,t} contains $u$. Let ${\cal G}_1 = (G_1,\lambda_1)$ be the temporal subgraph of ${\cal G}$ formed by the union of all \tpath{s,u}s, and define ${\cal G}_2 = (G_2,\lambda_2)$ similarly with respect to \tpath{u,t}s. We first make some observations concerning these temporal subgraphs.

    \begin{enumerate}[(I)]
        \item\label{item.Menger.I} $E^T({\cal G}_1)\cup E^T({\cal G}_2) = E^T({\cal G})$: this follows immediately from the fact that ${\cal G},s,t$ is a minimal counter-example. Indeed if $(e,i)$ is not contained in any \tpath{s,t}, then $tp_{{\cal G}-(e,i)}(s,t) = tp_{{\cal G}}(s,t)$ and $tpc_{{\cal G}-(e,i)}(s,t) = tpc_{{\cal G}}(s,t)$, contradicting the minimality of ${\cal G}, s,t$;

        \item\label{item.Menger.II} If there is an \tpath{s,u} passing by $(v,i)$ for some $v\neq u$, then there is no \tpath{u,t} passing by $(v,j)$, for every $j\ge i$: by contradiction, suppose that  $P_1$ is an \tpath{s,u} passing by $(v,i)$, and $P_2$ is a \tpath{u,t} passing by $(v,j)$, $j\ge i$. We get that $P' = sP_1(v,i)..(v,j)P_2t$ is an \twalk{s,t} not containing $u$, a contradiction as $P'$ contains an \tpath{s,t}, and by the choice of $u$. \\
              Observe that this property also implies that $E^T({\cal G}_1)\cap E^T({\cal G}_2) = \emptyset$.

        \item\label{item.Menger.III} $tpc_{{\cal G}_1}(s,u)>1$ and $tp_{{\cal G}_1}(s,u)>1$: first, observe that if $(v,j)\in V(G_1)\times  [\tau]$ is such that every \tpath{s,u} passes by $(v,j)$, since every \tpath{s,t} passes by $u$ we get that $(v,j)$ is also contained  in every \tpath{s,t}, a contradiction as $tpc_{\cal G}(s,t)>1$. Hence $tpc_{{\cal G}_1}(s,u)>1$.
              Now, observe that $E^T({\cal G}_1)\subset E^T({\cal G})$ since there is some \tpath{u,t} in ${\cal G}$ (i.e., $E^T({\cal G}_2)\neq \emptyset$) and by~\cref{item.Menger.II}.
              By the minimality of ${\cal G},s,t$ we get that $tp_{{\cal G}_1}(s,u) >1$;

        \item\label{item.Menger.IV} $tpc_{{\cal G}_2}(u,t)>1$ and $tp_{{\cal G}_2}(u,t)>1$: analogous to the previous item.
    \end{enumerate}

    To present our proof, we introduce some further definitions.
    For each $i\in \{1,2\}$, we denote the set of temporal edges of ${\cal G}_i$ incident to $u$ by $\delta_i$, i.e., $\delta_i = \delta^T_{{\cal G}_i}(u)$.
    Also, let $k_1 = tp_{{\cal G}_1}(s,u)$ and consider $k_1$ \tvertex-disjoint \tpath{s,u}s, $P_1,\ldots, P_{k_1}$. Let $S_1\subseteq \delta_1$
    be the set of temporal edges incident to $u$ contained in $\bigcup_{i=1}^{k_1}E^T(P_i)$; we say that $S_1$ is \emph{saturated by $P_1,\ldots,P_{k_1}$ in  ${\cal G}_1$}. We define a saturated set in $\delta_2$ similarly with regards to \tpath{u,t}s, and also let $k_2$ denote $tpc_{{\cal G}_2}(u,t)$.

    We first prove that, for every $(vu,i)\in \delta_1$, there exist $k_1$ \tvertex-disjoint \tpath{s,u}s $P_1,\ldots,P_{k_1}$ such that $(vu,i)$ is one of the temporal edges saturated by $P_1,\ldots,P_{k_1}$. For this, we consider arbitrary such paths and show that we can change them in order to saturate $(vu,i)$.
    Let $P$ be any \tpath{s,u} containing $(vu,i)$; it must exist by the definition of ${\cal G}_1$. Since $k_1 = tp_{{\cal G}_1}(s,u)$, we get that $P$ must intersect some path in $P_1,\ldots,P_{k_1}$. Let $(x,i')$ be a temporal vertex of $P$ closest to $u$ belonging to such an intersection, and suppose, without loss of generality, that $(x,i')$ appears in $P_1$. In other words, for every $j\in [k_1]\setminus\{1\}$, we have that $P' = (x,i')Pu$ intersects $P_j$ only in copies of $u$. First, suppose that there exists $(y,t_2)\in V^T(P')$ such that $y\notin \{x,u\}$ and $(y,t_1)\in V^T(P_j)$, for some $j\in[k_1]$ and $t_1\le t_2$. Suppose that $(y,t_2)$ is closest to $u$ in $P$ satisfying this choice, and that $P_j$ is chosen in order to maximize $t_1$.
    By the choice of $(x,i')$, we know that $t_1 < t_2$. Let $Q = sP_j(y,t_1)..(y,t_2)Pu$. We argue that $\{Q,P_1,\ldots,P_{k_1}\}\setminus \{P_j\}$ is a set of \tvertex-disjoint \tpath{s,u}s that saturate $(vu,i)$. Because $y\neq u$ and $P$ finishes by $(vu,i)$, we know that $Q$ saturates $(vu,i)$. Additionally, we know that $sP_j(y,t_1)$ does not intersect $P_h$ for every $h\neq j$, and by the choice of $(x,i')$, we know that $(y,t_2)Pu$ does not intersect $P_h$ for every $h\neq j$. Finally, by the choice of $P_j$ we know that $(y,t)\notin V^T(P_h)$ for every $h\neq j$ and every $t\in \{t_1+1,\ldots,t_2-1\}$ (the occurrences of $y$ within the waiting period). Note also that $Q$ is a path by the choice of $(y,t_2)$.

    Now suppose that no such \tvertex exists, i.e., that for every $(y,t_2)\in V^T(P')$ such that $y\notin \{x,u\}$, we have that if $(y,t_1)\in V^T(P_j)$ for some $j\in [k_1]$, then $t_1 > t_2$. Note that, in particular, this means that if $y\in V(sP_1(x,i'))$, for $y\neq x$, then $y\notin V(P')$; in words, if a vertex occurs in $P_1$ between $s$ and $(x,i')$, then $y$ does not occur in $P'$ between $(x,i')$ and $u$. Therefore $Q = sP_1(x,i)Pu$ is an \tpath{s,u}. Note also that, by the choice of $(x,i')$, we get that $\{Q,P_2,\ldots,P_{k_1}\}$ is a set of \tvertex-disjoint \tpath{s,u}s saturating $(vu,i)$.

    In what follows, it will be useful to combine an \tpath{s,u} $P_1$ with a \tpath{u,t} $P_2$ to form an \twalk{s,t}. This can be done whenever $P_1$ arrives in $u$ no later than $P_2$ departs from $u$; in this case we say that $P_1,P_2$ \emph{can be combined}. Observe that, since every temporal edge of $\mathcal{G}$ must be in some \tpath{s,t} (as we are in a minimal counter-example), and since every such path passes by $u$, we get that every \tpath{s,u} in ${\cal G}_1$ can be combined with \emph{some} \tpath{u,t} in ${\cal G}_2$, and vice-versa.

    Observe that the previous paragraphs actually give us that any set of $k_1$ \tvertex-disjoint \tpath{s,u}s can be modified by excluding one of them, and including a new one in order to saturate a given edge of $\delta_1$.
    We now use this fact to prove that $k_1 = \lvert \delta_1\rvert$. Clearly $\lvert \delta_1\rvert\ge k_1$ as any subset of $k_1$ \tvertex-disjoint \tpath{s,u}s must arrive in $u$ through distinct temporal edges.
    So, suppose by contradiction that $k_1 < \lvert \delta_1\rvert$, and consider $k_1$ arbitrary \tvertex-disjoint \tpath{s,u}s, $P_1,\cdots,P_{k_1}$. Also, consider $(vu,i)\in \delta_1$ not saturated by $P_1,\ldots,P_{k_1}$; it exists because $k_1 < \lvert \delta_1\rvert$. Denote $\mathcal{G}-(vu,i)$ by $\mathcal{G}'$.
    By \autoref{prop:remove_tempedge}, we know that there exists $(w,j)$ such that every \tpath{s,t} in ${\cal G}'$ contains $(w,j)$.  We argue that $w = u$ and that $P_h$ finishes in time at most $j$ for every $h\in [k_1]$.
    Recall that from~\cref{item.Menger.II} we get $E^T({\cal G}_1)\cap E^T({\cal G}_2) = \emptyset$; hence $(vu,i)\notin E^T({\cal G}_2)$. This and~\cref{item.Menger.IV} give us that $tpc_{{\cal G}'}(u,t)\ge 2$. This also implies that $(w,j)$ is a \tvcut{s,u} in ${\cal G}'$.
    Now observe that, by~\cref{item.Menger.III} and the choice of $(vu,i)$, all $k_1$ previously chosen paths exist in ${\cal G}'$. Hence, the only way for $(w,j)$ to break all possible combinations between $P_1,\ldots,P_{k_1}$ and the \tpath{u,t}s is if $w = u$ and $P_h$ finishes in time at most $j$ for every $h\in [k_1]$, as we wanted to argue.
    Finally, we can now apply the same argument to $k_1$ \tvertex-disjoint paths $P'_1,\ldots,P'_{k_1}$ obtained from the previous ones by replacing one of them with a path, say $P'_1$, that contains $(vu,i)$. Since $k_1\ge 2$, this implies that $i$ must also be at most $j$. However, this argument can be made for any arbitrary edge not saturated by $P_1,\cdots,P_{k_1}$, thus giving us that every $(xu,h)\in \delta_1$ is such that $h\le j$.  This is a contradiction since in this case $\{(u,j)\}$ is a \tvcut{s,t}.

    Finally, we have that $k_1 = \lvert \delta_1\rvert$. By analogous arguments we can conclude that $k_2 = \lvert \delta_2\rvert$.
    Note that if there exists $h\in  [\tau]$ such that $i\le h$ for every $(xu,i)\in \delta_1$, while $i\ge h$ for every $(ux,i)\in\delta_2$, then $\{(u,h)\}$ is a \tvcut{s,t} in $\mathcal{G}$, a contradiction; so suppose that this is not the case.
    Recall that every \tpath{s,u} in ${\cal G}_1$ can be combined with some \tpath{u,t} in ${\cal G}_2$, and vice-versa, to see that there must exist $(xu, i_1),(x'u, i'_1)\in \delta_1$, and $(uy,i_2),(uy',i'_2)\in \delta_2$ such that $i_1 \le i_2 < i'_1\le i'_2$. Finally note that, since $\delta_1$ and $\delta_2$ are saturated by maximum sets of \tvertex-disjoint  \tpath{s,u}s and \tpath{u,t}s, respectively, we get a contradiction since in this case we can obtain two \tvertex-disjoint \tpath{s,t}s by combining $P_1,P_2$ that pass by $(xu,i_1)$ and $(uy, i_2)$, respectively, and combining $P'_1,P'_2$ that pass by $(x'u,i'_1)$ and $(uy',i'_2)$, respectively.
\end{proof}

Now, given a minimal counter-example ${\cal G} = (G,\lambda),s,t$, let $S = \{(u,i_u),(v,i_v)\}$ be a \tvcut{s,t} in ${\cal G}$. By~\autoref{lem:nocutvertex}, we know that $u\neq v$. Below, we define the temporal subgraphs of interest.

\begin{itemize}
    \item ${\cal G}^S_{su} = (G_{su},\lambda_{su})$: union of all \tpath{s,u}s  not passing by $v$;
    \item ${\cal G}^S_{sv} = (G_{sv},\lambda_{sv})$: union of all \tpath{s,v}s not passing by $u$;
    \item ${\cal G}^S_{ut} = (G_{ut},\lambda_{ut})$: union of all \tpath{u,t}s not passing by $v$; and
    \item ${\cal G}^S_{vt} = (G_{vt},\lambda_{vt})$: union of all \tpath{v,t}s not passing by $u$.
\end{itemize}

For each $xy \in \{su,sv,ut,vt\}$ we denote by $T_S(x,y)$ the set of temporal vertices $(w,i)$ such that $(w,i)$ is contained in some path defining $\mathcal{G}^S_{xy}$.
In both notations, we omit $S$ when it is clear from the context.

Let ${\cal G}_1 = (G_1,\lambda_1)$ be equal to $(G_{su}\cup G_{sv},\lambda_{su}\cup \lambda_{sv})$, and ${\cal G}_2 = (G_2,\lambda_2)$ be equal to $(G_{ut}\cup G_{vt},\lambda_{ut}\cup \lambda_{vt})$. The general idea of the proof is to obtain paths from $s$ to $u$ and $v$ in ${\cal G}_1$ that can be combined with paths from $u$ and $v$ to $t$ in ${\cal G}_2$  in order to form two \tvertex-disjoint \tpath{s,t}s in $(G,\lambda)$, thus getting a contradiction.
This task is far from trivial as we need to ensure that $\mathcal{G}_1,\mathcal{G}_2$ are both smaller than $\mathcal{G}$ and that the paths in $\mathcal{G}_1$ arrive in $u,v$ in time to start the paths in $\mathcal{G}_2$.
We start by analyzing the easiest case, namely when $N(s)\cap N(t)\neq \emptyset$.

\begin{proposition}\label{prop:no_common_neighbor}
    Let ${\cal G} = (G,\lambda),s,t$ be a minimum counter-example. Then $N(s)\cap N(t)=\emptyset$.
\end{proposition}
\begin{proof}
    Suppose by contradiction that $u\in N(s)\cap N(t)$, and let $m = \min\lambda(su)$ and $M = \max\lambda(ut)$.
    By \autoref{lem:nocutvertex}, there exists an \tpath{s,t} $P$ in ${\cal G}$ not passing by $u$.
    Therefore, if $m\le M$, then $P$ and $(s,m,u,M,t)$ are two \tvertex-disjoint \tpath{s,t}s, a contradiction. We conclude that $m>M$. Now, let $P_u$ be an \tpath{s,t} containing $(su,m)$ and $P'_u$ be an \tpath{s,t} containing $(ut,M)$. Since $m > M$ and $(su,m)$ is the first temporal edge in $P_u$, observe that every temporal vertex of $P_u$ occurs at time bigger than $M$. But since $(ut,M)$ is the last temporal edge in $P'_u$, every temporal vertex of $P'_u$ occurs at time at most $M$. It follows that $P_u,P'_u$ are two \tvertex-disjoint \tpath{s,t}s, a contradiction.
\end{proof}

Now we prove a couple of useful propositions.

\begin{proposition}\label{prop:nonintersection}
    Let $(G,\lambda),s,t$ be a minimum counter-example, and $S = \{(u,i_u),(v,i_v)\}$ be a \tvcut{s,t} in $(G,\lambda)$. The following hold.
    \begin{enumerate}
        \item\label{item:GsuGut} $T(s,u)\cap T(u,t) \subseteq \{u\}\times  [\tau]$;
        \item\label{item:GsvGvt} Similarly, $T(s,v)\cap T(v,t) \subseteq \{v\}\times  [\tau]$;
        \item\label{item:GsuGvt} $T(s,u)\cap T(v,t) = \emptyset$; and
        \item\label{item:GsvGut} $T(s,v)\cap T(u,t) = \emptyset$.
    \end{enumerate}
\end{proposition}
\begin{proof}
    Suppose by contradiction that $(w,i)\in T(s,u)\cap T(u,t)$, where $w\neq u$, and let $P_1$ be an \tpath{s,u} in ${\cal G}_{su}$ containing $(w,i)$ and $P_2$ be a \tpath{u,t} in ${\cal G}_{ut}$ containing $(w,i)$. Define $P = sP_1(w,i)P_2t$.
    We argue that $P$ is an \twalk{s,t} not intersecting $\{u,v\}$, a contradiction as in this case we have that $S\cap V^T(P) = \emptyset$ and we know that $P$ contains an \tpath{s,t}. First note that $u\notin V(P)$ as $w\neq u$ and $u$ cannot be an internal vertex of neither $P_1$ nor $P_2$.
    Additionally, by the definition of ${\cal G}_{su}$ and ${\cal G}_{ut}$, we know that $v\notin V(P_1)\cup V(P_2)$. Observe that \cref{item:GsvGvt} is analogous.

    Now, suppose by contradiction that $(w,i)\in T(s,u)\cap T(v,t)$, and let $P_1$ be an \tpath{s,u} in ${\cal G}_{su}$ containing $(w,i)$ and $P_2$ be a \tpath{v,t} in ${\cal G}_{vt}$ containing $(w,i)$. Define $P = sP_1(w,i)P_2t$. As before, we argue that $P$ is an \twalk{s,t} not intersecting $\{u,v\}$, thus getting a contradiction. By the definition of ${\cal G}_{su}$ and ${\cal G}_{vt}$, we know that $v\notin V(P_1)$ and $u\notin V(P_2)$; therefore $w\notin \{u,v\}$. Our claim follows directly because we know that $u$ is not an internal vertex of $P_1$, nor $v$ an internal vertex of $P_2$. Observe that \cref{item:GsvGut} follows analogously.
\end{proof}

\begin{proposition}\label{prop:Gxy_non_empty}
    Let ${\cal G} = (G,\lambda),s,t$ be a minimum counter-example, and consider a \tvcut{s,t} in ${\cal G}$, $S = \{(u,i_u),(v,i_v)\}$. For each $x\in \{u,v\}$, there exist $P,P'$ such that $P$ is an \tpath{s,x} in ${\cal G}_{sx}$ arriving at time at most $i_x$, and $P'$ is an \tpath{x,t} in ${\cal G}_{xt}$ leaving at time at least $i_x$.
\end{proposition}
\begin{proof}
    Observe that there must exist an \tpath{s,t} $P$ containing $(u,i_u)$ such that $sPu$ does not contain $v$, as otherwise $v$ separates $s$ from $(u,i_u)$ and hence $\{v\}$ is a \tvcut{s,t}, contradicting \autoref{lem:nocutvertex}. The proposition follows by taking $sPu$. The other cases are analogous.
\end{proof}

Now, in order to get the desired paths in ${\cal G}_1$ and ${\cal G}_2$, we first need to ensure the existence of a convenient cut. The intuition behind the following definition is that we are searching for a cut $S = \{(u,i_u),(v,i_v)\}$ such that $(u,i_u)$ is the closest possible to $s$, while $(v,i_v)$ is the closest possible to $t$. Formally, given a \tvcut{s,t} $S = \{(u,i_u), (v,i_v)\}$, $i_u\le i_v$\change{, if $(x,i_x)\in T_S(s,u)\setminus \{(u,i_u)\}$ is such that $\{(x,i_x),(v,i_v)\}$ is a \tvcut{s,t}, then we say that $(x,i_x)$ is \emph{bad for $u$ in $S$ (with relation to $s$)}. Similarly, if $(y,i_y)\in T_S(v,t)\setminus \{(v,i_v)\}$ is such that $\{(u,i_u),(y,i_y)\}$ is a \tvcut{s,t}, then we say that $(y,i_y)$ is \emph{bad for $v$ in $S$} (with relation to $t$). We then say that $x\in \{u,v\}$ is \emph{good in $S$} if there are no bad vertices for $x$ in $S$. }

Observe that if $i_u=i_v$, then the roles of $u$ and $v$ can be swapped. \change{It will be clear from the text when we decide to do this. }
Finally, we say that $S$ is \emph{extreme} if $u$ and $v$ are good in $S$. We prove the existence of an extreme cut.
The following proposition helps us getting there.

\begin{proposition} \label{prop:otherCuts}
    Let ${\cal G} = (G,\lambda),s,t$ be a minimum counter-example and $S = \{(u,i_u),(v,i_v)\}$ be a \tvcut{s,t}. Let $(x,i_x)\in T_{\change{S}}(s,u)$ and $(y,i_y)\in T_{\change{S}}(v,t)$.  Then  \change{$(x,i_x)$ is bad for $u$ in $S$} if and only if every \tpath{s,u} in ${\cal G}^S_{su}$ arriving at time at most $i_u$ contains $(x,i_x)$. Similarly, \change{$(y,i_y)$ is bad for $v$ in $S$} if and only if every \tpath{v,t} in ${\cal G}^S_{vt}$ starting at time at least $i_v$ contains $(y,i_y)$.
\end{proposition}
\begin{proof}
    We first prove the necessary part of the first equivalence by contraposition. Let $P$ be an \tpath{s,u} in ${\cal G}^S_{su}$ arriving at time at most $i_u$ not containing $(x,i_x)$. By the definition of ${\cal G}^S_{su}$, we know that $P$ also does not contain $(v,i_v)$. Now, let $P'$ be a \tpath{u,t} in ${\cal G}^S_{ut}$ starting at time at least $i_u$; it exists by \autoref{prop:Gxy_non_empty}. Note that \autoref{prop:nonintersection} then gives us that $sPuP't$ is an \tpath{s,t} not intersecting $\{(x,i_x),(v,i_v)\}$.
    It follows that $\{(x,i_x),(v,i_v)\}$ is not a \tvcut{s,t}. Now suppose $(x,i_x)\in T_S(s,u)$ is such that $\{(x,i_x),(v,i_v)\}$ is not a \tvcut{s,t} in ${\cal G}$, and let $P$ be an \tpath{s,t} not passing by $\{(x,i_x),(v,i_v)\}$. Since $S$ is a cut, we get that $P$ must pass by $(u,i_u)$. If follows that $sPu$ is an \tpath{s,u} in ${\cal G}^S_{su}$ arriving at time at most $i_u$ not containing $(x,i_x)$, finishing the proof of the first equivalence.
    Observe that the second equivalence can be proved using similar arguments.
\end{proof}

Now, we show how to change a cut in order to get closer to an extreme cut.  Observe that if we apply this lemma to both vertices of the cut, then the obtained cut is extreme.

\begin{lemma}\label{lem:extremecut}
    Let ${\cal G} = (G,\lambda),s,t$ be a minimum counter-example and let $S = \{(u,i_u), (v,i_v)\}$ be any \tvcut{s,t}, $i_u\le i_v$. If $v$ is not good in $S$,  then there exists a cut $S' = \{(u,i_u), (y,i_y)\}$ where \change{$y$} is good in $S'$. Additionally, if $u$ is good in $S$, then $u$ is still good in $S'$. The analogous holds to replace $(u,i_u)$.
\end{lemma}
\begin{proof}
    \change{
        Let $\ell$ be maximum such that there exists some bad temporal vertex $(x,\ell)$ for $v$ in $S$. Also, let $B$ be the set of all such vertices, i.e., $B = \{x\in V(G)\mid (x,\ell)\text{ is bad for $v$ in $S$}\}$.
        For each $x\in B$, we denote by $S_x$ the set $\{(u,i_u),(x,\ell)\}$. Observe that, by definition of bad vertex, we know that $S_x$ is a \tvertex $s,t$-cut, for every $x\in B$. Additionally, \autoref{prop:Gxy_non_empty} tells us that there is at least one \tpath{x,t} in $\mathcal{G}^{S_x}_{xt}$ starting in time at least $\ell$; let $P_x$ be any such path.
        We first want to prove that there is a choice of $x\in B$ such that $x$ is good in $S_x$. For this, given two vertices $x,y\in B$, we write $x\prec y$ if $(y,\ell)$ is bad for $x$ in $S_x$.
        We prove that $\prec$ is anti-symmetric and transitive and then we argue that a maximal vertex $x$ in $\prec$ gives the desired temporal vertex.}

    \change{So, let $x,y\in B$ and suppose that $x\prec y$, which means that $(y,\ell)$ is bad for $x$ in $S_x$.}By \autoref{prop:otherCuts}, \change{we know that every $x,t$-path in $\mathcal{G}^{S_x}_{xt}$ starting in time at least $\ell$ contains $(y,\ell)$. In particular, $(y,\ell)$ is contained in $P_x$. We then have the \tpath{y,t}, $P' = (y,\ell)P_xt$, not containing $(x,\ell)$. Additionally, because $P$ is a path in $\mathcal{G}^{S_x}_{xt}$, we know that $P$ (and hence also $P'$) does not contain $u$. This means that $P'$ is a $y,t$-path contained in $\mathcal{G}^{S_y}_{yt}$ starting in time at least $\ell$ not containing $(x,\ell)$. Again}~\autoref{prop:otherCuts} \change{gives us that $x$ is not bad for $y$ in $S_y$. Therefore, $y\not\prec x$, as we wanted to prove.}

    \change{Now, to prove transitivity, consider $x,y,z,\in B$ such that $x\prec y$ and $y\prec z$. By definition and}~\autoref{prop:otherCuts}, \change{we know that every $x,t$-path in $\mathcal{G}^{S_x}_{xt}$ starting in time at least $\ell$ contains $(y,\ell)$, and that every $y,t$-path in $\mathcal{G}^{S_y}_{yt}$ starting in time at least $\ell$ contains $(z,\ell)$. In particular, $(y,\ell)$ is contained in $P_x$. As in the previous paragraph, we know that $P'=(y,\ell)Pt$ is a $y,t$-path contained in $\mathcal{G}^{S_y}_{yt}$ starting in time at least $\ell$. Therefore, $P'$ must contain $(z,\ell)$. Since this argument is applied to an arbitrary such path $P_x$, we get that $(z,\ell)$ is contained in every $x,t$-path in $\mathcal{G}^{S_x}_{xt}$ starting in time at least $\ell$. Again by}~\autoref{prop:otherCuts}\change{, we get that $x\prec z$, as we wanted to prove.}

    \change{Finally, let $x\in B$ be a $\prec$-maximal vertex. We argue that $(x,\ell)$ is good in $S_x$.  By contradiction, suppose that there exists $(y,i_y)\in T_S(x,t)\setminus \{(x,\ell)\}$ such that $\{(u,i_u),(y,i_y)\}$ is a \tvcut{s,t} (i.e., $(y,i_y)$ is bad for $x$ in $S_x$). By the definition of $\ell$ and since $x$ is $\prec$-maximal in $B$, we get that $i_y < \ell$. This contradicts}~\autoref{prop:otherCuts} \change{since $(y,i_y)$ cannot be within an $x,t$-path starting in time at least $\ell$.}

    \change{It now remains to prove that if $u$ is good in $S$, then $u$ is still good in $S_x$, where $x$ is chosen as in the previous paragraph.
        For this, let $\mathcal{G}^*_{su}$ be the temporal subgraph of $\mathcal{G}^S_{su}$ formed by the paths finishing in time at most $i_u$. Because $u$ is good in $S$, we know by}~\autoref{prop:otherCuts} \change{that there is no temporal vertex $(y,i_y)$ that is contained in every $s,u$-path in $\mathcal{G}^*_{su}$; in other words, we get $\tpc_{\mathcal{G}^*_{su}}(s,u)\ge 2$. We argue that every $s,u$-path contained in $\mathcal{G}^*_{su}$ is also contained in $\mathcal{G}^{S_x}_{su}$; note that this finishes the proof again by}~\autoref{prop:otherCuts}\change{. Let $P$ be an $s,u$-path in $\mathcal{G}^*_{su}$. If $P$ also does not contain $x$, then we are done, change{since $\mathcal{G}^{S_x}_{su}$ contains exactly the $s,u$-paths not containing $x$}. So suppose otherwise and let $P'$ be any \tpath{x,t} contained in $\mathcal{G}^S_{vt}$ starting in time at least $\ell$, which must exist by the choice of $x$ and by}~\autoref{prop:otherCuts}. \change{Observe now that, since $i_u\le i_v\le \ell$, we get that $sPxP't$ is an $s,t$-walk not intersecting $\{u,v\}$. Since any walk contains a path, we get a contradiction to the fact that $S$ is a \tvcut{s,t}. Therefore this case does not occur and indeed $P$ is also contained $\mathcal{G}^{S_x}_{su}$, as we wanted to prove.}
\end{proof}

In our final step, we want to be able to apply induction on some temporal graphs smaller than ${\cal G}$, and for this we need further definitions.
We say that $S$ is \emph{$t$-dominated} if ${\cal G}_2$ is formed just by the temporal edges $(ut,\ell_u)$ and $(vt,\ell_v)$, for some $\ell_u,\ell_v$. We define $s$-dominated similarly, i.e., $S$ is \emph{$s$-dominated} if ${\cal G}_1$ is formed just by the temporal edges $(su,\ell_u)$ and $(sv,\ell_v)$, for some $\ell_u,\ell_v$. Finally, we say that $S$ is \emph{dominated} if it is either $t$-dominated or $s$-dominated.

\begin{lemma}\label{lem:nondominated_cut}
    If Item~1 of \autoref{thm:menger} does not hold, then there exists a minimum counter-example ${\cal G} = (G,\lambda),s,t$ that has a non-dominated extreme \tvcut{s,t}.
\end{lemma}
\begin{proof}
    Let ${\cal G} = (G,\lambda),s,t$ be a counter-example minimizing $|V(G)|+|E^T({\cal G})|$. If no cut is dominated, then we can apply~\autoref{lem:extremecut} on any cut to obtain an extreme cut.
    So suppose that $\delta^T(s) = \{(su,i_u),(sv,i_v)\}$, $i_u\le i_v$; the case where $\lvert \delta^T(t)\rvert = 2$ is analogous.
    Let $S = \{(u,i_u),(v,i_v)\}$, and note that $S$ is a \tvcut{s,t}.
    \change{Observe that, since $(su,i_u)\in E^T(\mathcal{G})$, there cannot be any temporal vertex $(x,i_x)$ contained in every $s,u$-path arriving in time at most $i_u$. By \autoref{prop:otherCuts} we then get that $u$ is good in $S$ with relation to $s$. Note also that, in case $i_u=i_v$, then the same argument can be applied to $v$, i.e., $v$ is good in $S$ with relation to $s$. This means that the roles of $u$ and $v$ can be switched in case $i_u=i_v$.
        We now prove the following nice property.

        \medskip
        {\bf(*)} If $v$ is not good in $S$ with relation to $t$, then the lemma follows.

        \medskip
        To prove (*), let $S' = \{(u,i_u),(x,i_x)\}$ be an extreme cut obtained by applying \autoref{lem:extremecut}. We need to prove that $S'$ is not dominated. By \autoref{prop:no_common_neighbor}, we know that $u\notin N(t)$; hence $S'$ is not $t$-dominated. Also, recall that $(su,i_u),(sv,i_v)$ are the only temporal edges incident to $s$ (by definition of $s$-dominated) and, since $(x,i_x)\neq (v,i_v)$, we also get that $S'$ is not $s$-dominated, as we wanted to prove.}

    \change{The general idea of the proof is to show first that we can suppose $i_u=i_v$, then prove that either $u$ or $v$ is not good in $S$ with relation to $t$. Since $u$ and $v$ are interchangeable because $i_u=i_v$, the lemma follows by Property~(*).}

    \change{As previously said,} we first prove that we can suppose $i_u=i_v$.
    Note that, since $(su,i_u)$ is the first edge incident to $s$, we \change{can suppose that} $E^T({\cal G})\subseteq V(G)\times\{i_u,\cdots,\tau\}$, which in particular means that there are no edges incident to $u$ before timestep $i_u$.
    We now argue that either there are no edges incident to $v$ before timestep $i_v$, or  \change{Property~(*) holds and the lemma follows}.
    Suppose that $(vx,i_x)\in E^T({\cal G})$ is such that $i_x < i_v$. Let ${\cal G}'$ be obtained from ${\cal G}$ by removing $(vx,i_x)$. By \autoref{prop:remove_tempedge}, we know that there exists a \tvcut{s,t} of size~1, $\{(y,i_y)\}$, in ${\cal G}'$. Note that if there exists a \tpath{v,t} $P$ starting at time at least $i_v$ not passing by $(y,i_y)$, then $P$ also cannot contain $(xv,i_x)$ as $i_x < i_v$. Hence,
    $s(v,i_v)Pt$ is an \tpath{s,t} in ${\cal G}'$ not passing by $(y,i_y)$, a contradiction.
    Therefore $(y,i_y)$ is within every \tpath{v,t} in ${\cal G}$ starting at time at least $i_v$, and by \autoref{prop:otherCuts}, we get that $v$ is not good in $S$ \change{with relation to $t$. The lemma follows by Property~(*)}.

    Now, we argue that if $\mathcal{G}'$ is obtained from $\mathcal{G}$ by relabeling $sv$ with $i_u$, then $tp_{\mathcal{G}'}(s,t)= tp_{\mathcal{G}}(s,t)$ and $tpc_{\mathcal{G}'}(s,t)= tpc_{\mathcal{G}}(s,t)$. \change{Observe that this finishes the part of the proof where we wanted to assume $i_u=i_v$}. Clearly, if $P$ is an \tpath{s,t} in $\mathcal{G}$ containing $sv$, then $P$ is still a path in $\mathcal{G}'$  and vice-versa, because $i_u\le i_v$ and by what is proved in the previous paragraph. This is also why the size of a cut cannot decrease.

    \change{To finish the proof, we want to show that either $u$ or $v$ is not good in $S$ with relation to $t$; so suppose otherwise, i.e., that both $u$ and $v$ are good in $S$ with relation to $t$. Let $(yx,f)$ be the first temporal edge incident to either $u$ or $v$ and not incident to $s$. More formally, $(yx,f) \in (\delta^T(u)\cup \delta^T(v))\setminus\delta^T(s)$ minimizes $f$.
        Because we can switch the roles of $u$ and $v$, we can suppose, without loss of generality, that $y=u$.}
    Observe that we can also suppose that $f=i$ as otherwise we can relabel $su$ and $sv$ with $f$.
    Let ${\cal G}' = {\cal G}-\{(ux,i)\}$ and let $\{(y,i_y)\}$ be a \tvcut{s,t} in ${\cal G}'$ ensured by \autoref{prop:remove_tempedge}. We first argue that $y\neq v$.
    Let $\mathcal{G}^*_{ut}$ contain every \tpath{u,t} in $\mathcal{G}_{ut}$ starting in time at least $i$. Because $u$ is good in $S$ with relation to $t$, \autoref{prop:otherCuts} \change{gives us that no temporal vertex is contained in every \tpath{u,t} in $\mathcal{G}_{ut}$ starting in time at least $i$}. In other words,
    $tpc_{\mathcal{G}^*_{ut}}(u,t)\ge 2$. And since $\mathcal{G}^*_{ut}$ is smaller than $\mathcal{G}$ \change{and $\mathcal{G}$ is a minimum counter-example, we know that $\mathcal{G}^*_{ut}$ cannot be a counter-example, i.e., it cannot happen that $tpc_{\mathcal{G}^*_{ut}}(u,t)\ge 2$ while $tp_{\mathcal{G}^*_{ut}}(u,t)=1$}. We then get $tp_{\mathcal{G}^*_{ut}}(u,t)\ge 2$.
    Therefore there exists a \tpath{u,t} in ${\cal G}_{ut}$ not passing by $(ux,i)$, which together with $(su,i)$ gives us an \tpath{s,t} in ${\cal G}'$ not passing by $v$. It follows that $y\neq v$, as we wanted to prove.

    Now, observe that $S' = \{(u,i),(y,i_y)\}$ is a \tvcut{s,t} in $\mathcal{G}$ as indeed every \tpath{s,t} in $\mathcal{G}$ either passes by $(y,i_y)$ or by $(ux,i)$ and hence by $(u,i)$. We argue that $(y,i_y)$ is within every \tpath{v,t} in $\mathcal{G}_{vt}$ starting in time at least $i$. Note that this finishes the proof as it implies that $v$ is not good in $S$ with relation to $t$. Suppose otherwise and let $P$ be a \tpath{v,t} in  $\mathcal{G}_{vt}$ starting in time at least $i$ not passing by $(y,i_y)$. By the definition of $\mathcal{G}_{vt}$, we know that $P$ also does not contain $u$ and therefore does not pass by $(ux,i)$. But then concatenating $(sv,i)$ with $P$ we get an \tpath{s,t} in $\mathcal{G}'$ not containing $(y,i_y)$, contradiction.
\end{proof}

Recall that ultimately we want to combine paths from ${\cal G}_1$ with paths from ${\cal G}_2$, but in order to do that we need to ensure that they can be combined, i.e., that the ones taken in ${\cal G}_1$ arrive in $u$ and $v$ in time to continue through the ones taken in ${\cal G}_2$. This idea is captured in the following definition.
\change{We say that $u$ is \emph{nice in $S$} if $j\le i_u$ for every $(xu,j)\in \delta^T_{{\cal G}_{su}}(u)$ and $j\ge i_u$ for every $(ux,j)\in \delta^T_{{\cal G}_{ut}}(u)$. We define the notion of $v$ being nice in $S$ in the same way, just replacing $u$ by $v$ previously.}

\begin{lemma}\label{lem:uv_nice}
    Let ${\cal G} = (G,\lambda),s,t$ be a minimal counter-example, and $S=\{(u,i_u),(v,i_v)\}$ be an extreme \tvcut{s,t} in ${\cal G}$, where $i_u\le i_v$. Then $u$ and $v$ are nice in $S$.
\end{lemma}
\begin{proof}
    We prove that, for each $x\in \{u,v\}$ and $i$ such that $(yx,i)\in E^T({\cal G}_{sx})$, we have that $i\le i_x$.
    The proof that $i\ge i_x$ for each $x\in \{u,v\}$ and $i$ such that $(xy,i)\in E^T({\cal G}_{xt})$ is analogous. \change{The main difference is that one needs to flip also the order of the proof. The first part of the proof below, that concerns the non-existence of $(yv,i)\in E^T(\mathcal{G}_{sv})$ such that $i>i_v$, can be shadowed to get a proof that there is no $(yu,i)\in E^T(\mathcal{G}_{ut})$ such that $i<i_u$. Similarly, the second part can be shadowed to prove that there is no $(yv,i)\in E^T(\mathcal{G}_{vt})$ such that $i<i_v$. In general, it is enough to switch the roles of $u$ and $v$, and the roles of $s$ and $t$.}

    First suppose that there exists $(yv,i)\in E^T({\cal G}_{sv})$ such that $i>i_v$, and let $P$ be an \tpath{s,t} passing by $(yv, i)$.
    \change{We analyze two cases.}

    \begin{itemize}
        \item \change{$P$ contains $(v,i_v)$: this means that $P$ arrives in $v$ at a time $a\le i_v$ and leaves from $v$ at time $i$ through the temporal edge $(vy,i)$. If $u\notin V(vPt)$, then $vPt$ is a path in $\mathcal{G}_{vt}$. This contradicts \cref{item:GsvGvt} of \autoref{prop:nonintersection} as in this case $(y,i)$ would be contained in $T(s,v)\cap T(v,t)$. We can then conclude that $u\in V(vPt)$. Additionally, observe that, since $i_u\le i_v < i$ and $u$ appears after $(y,i)$ in $P$, we get that $(u,i_u)\notin V^T(vPt)$. Now, let $P'$ be an $s,t$-path in $\mathcal{G}_{sv}$ passing by $(yv,i)$ (it exists since $(yv,i)\in E^T({\cal G}_{sv})$). Because $P'$ arrives at $v$ at time $i>i_v$ and $P'$ is a path, we know that $P'$ does not contain $(v,i_v)$. By definition of $\mathcal{G}_{sv}$, we know that $P'$ also does not contain $u$. We then get that $sP'vPt$ is an $s,t$-path not intersecting $S$, a contradiction.}

        \item $P$ does not contain $(v,i_v)$: since $S$ is a \tvcut{s,t}, it follows that $P$ must contain $(u,i_u)$. We argue that $(u,i_u)$ must occur after $(yv,i)$ in $P$; this gives us a contradiction as in this case we would have $i_v < i \le i_u$. Suppose otherwise, i.e., that $(u,i_u)$ occurs before $(yv,i)$ in $P$. Note that if $(y,i)$ occurs after $(v,i)$ in $P$, then $vPt$ is a \tpath{v,t} not containing $(u,i_u)$ and containing $(y,i)$, i.e., $(y,i)\in  T(v,t)$. But since $(y,i)\in T(s,v)$ and $y\neq v$ by the choice of the temporal edge $(yv,i)$, we get a contradiction to \change{\cref{item:GsvGvt}} of \autoref{prop:nonintersection}. Therefore $(y,i)$ occurs before $(v,i)$ in $P$.
              Now, let $P'$ be an \tpath{s,v} in ${\cal G}_{sv}$ containing $(yv,i)$ (it exists since $(yv,i)\in E^T({\cal G}_{sv})$). Because $i>i_v$, \change{$P'$ contains only $(v,i)$ as it is a path}, and by the definition of ${\cal G}_{sv}$ \change{(it is formed by paths not containing $u$)}, we know that $P'$ does not intersect $S$. Similarly, because $P$ arrives in $v$ in time $i>i_v$ and $(u,i_u)$ occurs before $v$ in $P$, we get that $(v,i)Pt$ also does not intersect $S$. Therefore $P'' = sP'(v,i)Pt$ is an $s,t$-walk in ${\cal G}$ not intersecting $S$, and observe that the \tpath{s,t} contained in $P''$ also does not intersect $S$, contradiction.
    \end{itemize}

    \change{Observe that in the above proof we use the fact that $i_u\le i_v$; hence we need a different proof for $u$. }
    Suppose that there exists $(yu,i)\in E^T({\cal G}_{su})$ such that $i>i_u$, and
    let ${\cal G}'$ be obtained from ${\cal G}$ by removing $(yu,i)$. Note that  \change{$y\neq v$} by the definition of $\mathcal{G}_{su}$ \change{and that $y\neq u$} since $yu\in E(\mathcal{G})$ \change{and $\mathcal{G}$ does not contain loops}.
    By \autoref{prop:remove_tempedge}, there exists a \tvcut{s,t} in ${\cal G}'$ of size~1, say $\{(x,i_x)\}$.
    First we \change{prove the following claims.}

    \begin{claim}\label{claim1:lemma5}
        $S ' = \{(x,i_x),(v,i_v)\}$ is a \tvcut{s,t} in ${\cal G}$
    \end{claim}
    \begin{proof}[Proof of claim]
        By contradiction, let $P$ be an \tpath{s,t} in ${\cal G}$ not intersecting $S'$.
        \change{Because $(x,i_x)\notin V^T(P)$ and $\{(x,i_x)\}$ is a \tvertex $s,t$-cut in $\mathcal{G}'$, we know that} $P$ does not exist in ${\cal G}'$. \change{This in turn gives us that} $P$ must contain $(yu,i)$.
        \change{Observe that $P$ must contain $(u,i_u)$ as it does not contain $(v,i_v)$. This means that $P$ arrives in $u$ at a time $a\le i_u$ and leaves from $u$ at time $i>i_u$ through the edge $(uy,i)$. Now, let $P'$ be an $s,t$-path in $\mathcal{G}_{su}$ passing by $(yu,i)$ (it exists since $(yu,i)\in E^T(\mathcal{G}_{su}$)). By the definition of $\mathcal{G}_{su}$, we know that $P'$ does not contain $v$. Additionally, as $P'$ is a path and it arrives too late in $u$ in order to contain $(u,i_u)$, we know that $P'$ does not intersect $S$. We then get that $sP'uPt$ is an $s,t$-path not intersecting $S$, a contradiction.}
    \end{proof}

    \begin{claim}\label{claim2:lemma}
        $(x,i_x)\notin T(s,v)$.
    \end{claim}
    \begin{proof}[Proof of claim]
        \change{We actually argue that $(x,i_x)\in T(u,t)$.
            The claim then follows by \cref{item:GsvGut} of \autoref{prop:nonintersection}}.
        \change{Now, observe that if $(x,i_x)\in T(s,u)$, by \autoref{claim1:lemma5} we get that $(x,i_x)$ is bad for $u$ in $S$, contradicting the fact that}
        $S$ is extreme. We must then have that $(x,i_x)\notin  T(s,u)$.  By  \autoref{prop:Gxy_non_empty}\change{, there exists an $s,u$-path $P$ in $\mathcal{G}_{su}$ arriving at time at most $i_u$ not containing $v$ and a $u,t$-path in $\mathcal{G}_{ut}$ starting at time at least $i_u$ not containing $v$. Therefore $P'' = sPuP't$ contains an $s,t$-path not containing $v$.} \change{By \autoref{claim1:lemma5}, we then get that $P''$ must contain $(x,i_x)$. But since $(x,i_x)\notin T(s,u)$ (and hence not in $P$), we get that $(x,i_x)$ occurs in $P'$, and hence}  $(x,i_x)\in T(u,t)$, as we wanted to prove.
    \end{proof}

    We finally argue that $\{(x,i_x)\}$ must be a cut in ${\cal G}$, a contradiction \change{to \autoref{lem:nocutvertex}}.
    Suppose otherwise, and let $P$ be an \tpath{s,t} in ${\cal G}$ not passing by $(x,i_x)$.
    Then it must contain $(yu,i)$ as $\{(x,i_x)\}$ is a \tvcut{s,t} in ${\cal G}'$. \change{We now prove that $(v,i_v)$ must occur in $P$ after $(uy,i)$. For this, we analyze two cases.}

    \begin{itemize}
        \item \change{$P$ contains $(u,i_u)$: this means that $P$ arrives in $u$ at a time $a\le i_u$ and leaves from $u$ at time $i$ through the temporal edge $(uy,i)$. If $v\notin V(uPt)$, then $uPt$ is a path in $\mathcal{G}_{ut}$. This contradicts \cref{item:GsuGut} of \autoref{prop:nonintersection} as in this case $(y,i)$ would be contained in $T(s,u)\cap T(u,t)$. We can then conclude that $v\in V(uPt)$. Now suppose that $(v,i_v)\notin V^T(uPt)$.
                  Let $P'$ be an \tpath{s,u} in $\mathcal{G}_{su}$ passing by $(yu,i)$(it exists since $(yu,i)\in E^T(\mathcal{G}_{su})$). By the definition of $\mathcal{G}_{su}$ and since $P'$ is a path, we get that $V(sP'y)\cap \{u,v\}=\emptyset$. Additionally, we know that $u\notin V(yPt)$; hence $P^*=sP'yPt$ is an $s,t$-walk not containing $u$. Observe also that, because $(v,i_v)\notin V^T(P)$ and $v\notin V(P')$, we get that the \tpath{s,t} contained in $P^*$ also does not contain $(v,i_v)$, a contradiction as $S$ is a \tvcut{s,t}.}

        \item $P$ does not contain $(u,i_u)$: since $S$ is a \tvcut{s,t} in $\mathcal{G}$, we get that $P$ must contain $(v,i_v)$.
              Suppose  \change{$(v,i_v)$ occurs before $(yu,i)$ in $P$} and observe that, if $y$ occurs after $u$ in $P$, then $uPt$ is a \tpath{u,t} not containing $v$, which implies $(y,i)\in T(u,t)$. This contradicts \change{\cref{item:GsuGut} of} \autoref{prop:nonintersection} as we also have $(y,i)\in T(s,u)$. It follows that $u$ occurs after $y$ in $P$. We can apply an argument similar to previous ones to get a contradiction \change{to the fact that $S$ is a \tvcut{s,t}} by concatenating an \tpath{s,u} passing by $(yu,i)$ with $(u,i)Pt$. Therefore, we conclude that $(v,i_v)$ occurs in $P$ after the appearance of $(yu,i)$, as we wanted.
    \end{itemize}

    Finally, let $P'$ be an \tpath{s,v} in ${\cal G}_{sv}$ arriving at time $j\le i_v$  (it exists by \autoref{prop:Gxy_non_empty}).  We get that $P^* = sP'vPt$ is an $s,t$-walk in ${\cal G}$.
    In fact, \change{\cref{item:GsvGvt} of} \autoref{prop:nonintersection} and the fact that $P'$ is a path in ${\cal G}_{sv}$ and $(v,i_v)Pt$ is a path in ${\cal G}_{vt}$ give us that $P^*$ is actually an $s,t$-\emph{path} in ${\cal G}$.
    \change{To finish the proof, we argue that $P^*$ does not contain $(yu,i)$ nor $(x,i_x)$, getting a contradiction as $\{(x,i_x)\}$ is a \tvcut{s,t} in $\mathcal{G}'$. By \autoref{claim2:lemma}, the fact that $P'$ is a path in $\mathcal{G}_{sv}$, and by the choice of $P$, we know that $(x,i_x)\notin V^T(P^*)$. Additionally, we know that $(yu,i)\notin E^T(vPt)$ because $(v,i_v)$ occurs after it in $P$. Finally, $P'$ also cannot contain $(yu,i)$ because it is a path in $\mathcal{G}_{sv}$ (i.e., it does not contain $u$).}
\end{proof}

We are finally ready to finish the proof of \autoref{thm:menger}.

\begin{proof}[Final steps of the proof of Item~1 of \autoref{thm:menger}]
    By contradiction, let ${\cal G}, s, t$ be a minimum counter-example for Item~1 of \autoref{thm:menger}.
    By \autoref{lem:nondominated_cut}, we can suppose that there exists an extreme non-dominated \tvcut{s,t} in ${\cal G}$, $S = \{(u,i_u),(v,i_v)\}$. By \autoref{lem:uv_nice}, we also get that $u$ and $v$ are nice in $S$. We now construct two \tvertex-disjoint \tpath{s,t}s in ${\cal G}$, thus getting a contradiction.

    Recall that ${\cal G}_1 = (G_1,\lambda_1)$ is equal to $(G_{su}\cup G_{sv},\lambda_{su}\cup \lambda_{sv})$.
    Let ${\cal G}'_1 = (G'_1,\lambda'_1)$ be obtained from ${\cal G}_1$ by adding a new vertex $t'$ adjacent to $u$ and $v$, and making $\lambda'_1(ut') = \lambda'_1(vt') = \{\tau\}$.
    Note that, by \autoref{prop:nonintersection}, we have that $E^T({\cal G}_1)\cap E^T({\cal G}_2) = \emptyset$. Hence, since $S$ is not $t$-dominated, we get that $|V(G'_1)|+|E^T({\cal G}'_1)| <  |V(G)|+|E^T({\cal G})|$, and therefore, ${\cal G}'_1,s,t'$ is not a counter-example.
    First observe that, $tp_{{\cal G}'_1}(s,t')\ge 1$ by \autoref{prop:Gxy_non_empty}; hence $tpc_{{\cal G}'_1}(s,t')\ge 1$.
    Now suppose by contradiction that $tpc_{{\cal G}'_1}(s,t')=1$, and let $(x,i)$ be a \tvcut{s,t'} in ${\cal G}'_1$. Because $x\neq t'$, we get that $(x,i)\in T({\cal G})$; and since $\{(x,i)\}$ is not a \tvcut{s,t} in ${\cal G}$, there exists an \tpath{s,t} $P$ in ${\cal G}$ not containing $(x,i)$. Suppose, without loss of generality, that $u$ appears before $v$ in $P$ and let $j$ be the time of arrival of $P$ in $u$; hence $sP(u,j)..(u,\tau)t'$ is an \tpath{s,t'} in ${\cal G}'_1$ not containing $\{(x,i)\}$, a contradiction.
    We therefore get that $tpc_{{\cal G}'_1}(s,t')\ge 2$. Again because ${\cal G}'_1,s,t'$ is not a counter-example, we must have $tp_{{\cal G}'_1}(s,t')\ge 2$. Note that this implies the existence of paths $P_u,P_v$ such that $P_u$ is an \tpath{s,u} in ${\cal G}_1$, $P_v$ is an \tpath{s,v} in ${\cal G}_1$, and $P_u,P_v$ intersect only in $s$. Observe also that a similar argument can be applied to obtain paths $P'_u,P'_v$ such that $P'_u$ is a \tpath{u,t} in ${\cal G}_2$, $P'_v$ is an \tpath{v,t} in ${\cal G}_2$, and $P'_u,P'_v$ intersect only in $t$. Finally, since $u$ and $v$ are nice in $S$, and by \autoref{prop:nonintersection}, we get that $sP_uP'_ut$ and $sP_vP'_vt$ contain the desired two \tvertex-disjoint \tpath{s,t}s in ${\cal G}$, as we wanted to prove.
\end{proof}

In \autoref{sec:paths_algorithm}, we show how to find 2 \tvertex-disjoint \tpath{s,t}, if they exist. However, we first need to find cuts of fixed sizes. This is done in the following subsection.


\subsection{Parameterized algorithms for cuts}\label{sec:XP}

Given a temporal graph ${\cal G} =(G,\lambda)$, non-adjacent vertices $s,t\in V(G)$, and a subset $S\subseteq (V(G)\setminus\{s,t\})\times [\tau]$, consider the problem of deciding whether a given set $S$ is a \tvcut{s,t}. By \autoref{thm:decide_if_cut}, we know that such problem is co-$\NP$-complete.
In this section, we parameterize this problem by the size of $S$ and we give an $\XP$ algorithm to solve it. As we argue in more details later on, this immediately gives an $\XP$ algorithm for {\htcuts}.
The reader should observe that the arguments apply also when ${\cal G}$ is a temporal directed graph. Additionally, as this is a positive result, we get that it also applies to the strict context by \autoref{thm:strict_to_nonstrict}.

\begin{theorem}\label{thm:is_S_cut_inXP}
    Given a temporal (directed) graph ${\cal G} = (G,\lambda)$ with lifetime $\tau$, non-adjacent vertices $s,t\in V(G)$, and $S\subseteq (V(G)\setminus \{s,t\})\times [\tau]$ of size at most $h$,  deciding whether $S$ is a \tvcut{s,t} in ${\cal G}$ can be done in time $O^*(h^h)$.
\end{theorem}
\begin{proof}
    For each $u\in V(G)\setminus \{s,t\}$ appearing in $S$, denote by $\mathcal{L}_u=\{L_0,\cdots,L_{t_u}\}$ the interval windows defined by the appearances of $u$ in $S$. More formally, suppose that $\{(u,i_1),\ldots,(u,i_{t_u})\}$ are all the appearances of $u$ in $S$, with $i_1 < \ldots < i_{t_u}$. Then $L_0$ is equal to the interval $[1,i_1)$, $L_{t_u}$ is equal to the interval $(i_{t_u},\tau]$, and $L_j$ is equal to the interval $(i_j,i_{j+1})$.
    Now, let $U = \{u_1,\ldots,u_q\}\subseteq V(G)$ be the set of all vertices that appear at least once in $S$; denote $t_{u_i}$ by $t_i$ for each $u_i\in U$. For every choice of values $I=\{j_1,\ldots,j_q\}$ where $j_i\in \{0,\ldots,t_{i}\}$ for each $i\in[q]$, let ${\cal G}_I$ be the temporal graph obtained from ${\cal G}$ by removing all the temporal edges incident in $u_i$ in a timestamp not in $L_{j_i}$ for every $i\in [q]$. More formally, ${\cal G}_I = (G,\lambda_I)$, where
    \[\lambda_I(uv) = \left\{\begin{array}{ll}
            \lambda(uv)                         & \mbox{, if $u,v\notin U$, }                     \\
            \lambda(uv)\cap L_{j_i}             & \mbox{, if $u = u_i\in U$ and $v\notin U$, and} \\
            \lambda(uv)\cap L_{j_i}\cap L_{j_h} & \mbox{, if $u = u_i\in U$ and $v = u_h\in U$}.
        \end{array}\right.\]

    We prove that $S$ is not a \tvcut{s,t} in ${\cal G}$ if and only if there exists a choice of $I$ for which ${\cal G}_I$ contains an \twalk{s,t}. The problem therefore reduces to testing, for every choice of $I$, whether there exists an \twalk{s,t} in ${\cal G}_I$, which can be done in time $O(m\tau)$ using the algorithm proposed in~\cite{Wu.etal.16}.
    Observe that there are $\prod_{i=1}^q (t_i+1)$ possible choices of $I$. Since $q\le h$ and $t_i\le h$ for each $i\in[q]$, it follows that $\prod_{i=1}^q (t_i+1) \le (h+1)^h = O(h^h)$.

    So now suppose that $S$ is not a \tvcut{s,t} in ${\cal G}$, and let $P$ be an \tpath{s,t} in ${\cal G}$ not intersecting $S$. For each $u_i\in V(P)\cap U$, let $\ell^1_i,\ell^2_i$ be the timesteps of the temporal edges of $P$ incident to $u_i$ with $\ell^1_i\le \ell^2_i$. We know that there are exactly 2 such edges since $P$ is a path and $u_i\notin \{s,t\}$. Also, because $P$ does not intersect $S$, we get that $(u_i,t)\notin S$ for every $t\in [\ell^1_i,\ell^2_i]$. This means that $[\ell^1_i,\ell^2_i]\subseteq L_{j_i}$ for some $j_i\in \{0,\ldots,t_i\}$. By choosing such $j_i$ for every $u_i\in V(P)\cap U$, and letting $j_i = 0$ for every $u_i\in U\setminus V(P)$, we get that $P$ is contained in ${\cal G}_I$, where $I = \{j_1,\ldots, j_q\}$, as we wanted to show.

    Finally, suppose that there exists $I = \{j_1,\ldots,j_q\}$, and an \twalk{s,t} $W$  in ${\cal G}_I$. Observe that, because ${\cal G}_I\subseteq {\cal G}$, the \tpath{s,t} $P$ contained in $W$ is a path in ${\cal G}$. It remains to argue that $P$ does not intersect $S$, which implies that $S$ is not a \tvcut{s,t} in ${\cal G}$. Suppose otherwise and let $(u_i,t)\in S\cap V^T(P)$; also let $(e_1,\ell_1)$ and $(e_2,\ell_2)$ be the two temporal edges of $P$ incident to $u_i$. Observe that $t$ is not within any of the defined intervals for $u_i$, which means that, regardless of the choice of $j_i$, there are no temporal edges in ${\cal G}_I$ incident in $(u_i,t)$; hence $\ell_1 < t < \ell_2$. Note also that the interval chosen for $u_i$ is either to the left or to the right of $(u_i,t)$, i.e., $L_{j_i}$ is either contained in $[1,t)$ or in $(t,\tau]$. This is a contradiction as in this case either $\ell_1\notin L_{j_i}$ ( and hence $(e_1,\ell_1)\notin E^T({\cal G}_I)$) or  $\ell_2\notin L_{j_i}$ (and hence $(e_2,\ell_2)\notin E^T({\cal G}_I)$).
\end{proof}

As a consequence, we get the following:

\begin{corollary}
    \htcuts\ can be solved in time~$O((hn\tau)^h)$, where $n = \lvert V(G)\rvert$.
    \label{cor:cuts_XP}
\end{corollary}
\begin{proof}
    We can simply test, for every choice of $S\subseteq (V(G)\setminus\{u,v\})\times [\tau]$ of size $h$, whether $S$ is a \tvcut{s,t}. Since there are $n\tau$ possible temporal vertices, we have $O((n\tau)^h)$ possible choices. For each such choice, by \autoref{thm:is_S_cut_inXP}, we take $O^*(h^h)$ time to do the test. The corollary follows directly.
\end{proof}


\subsection{Algorithm for finding two \tvertex-disjoint paths}\label{sec:paths_algorithm}

An interesting byproduct of \autoref{thm:menger} and \autoref{cor:cuts_XP} is that they help us solving {\ktpaths} for the next fixed value $k=2$. In fact, we can not only solve the decision problem, but also find the desired paths. Again, we work on the non-strict problem and apply~\autoref{thm:strict_to_nonstrict} if we want to solve the strict case. We first need some tool lemmas.

\begin{lemma}
    Let $\mathcal{G}$ be a temporal (directed) graph and consider a pair of non-adjacent vertices. Then, $\tpc_{{\cal G}}(s,t) \geq 2$ if and only if $\tp_{{\cal G}}(s,t) \geq 2$. Additionally,  we can decide whether $\tp_{{\cal G}}(s,t) \geq 2$ in time $O((m+n)\tau)$.
    \label{lem:tpc-2-iff-tp-at-least-2}
\end{lemma}
\begin{proof}
    By \autoref{thm:menger}, we know that $\tp(s,t) \le 1$ if and only if $\tpc(s,t)\le 1$. By contrapositive, if $\tpc(s,t)\ge 2$, then $\tp(s,t)\ge 2$. Additionally, if $\tp(s,t)\ge 2$, then $\tpc(s,t)\ge 2$ follows directly as any cut must intersect each path of a set of \tvertex-disjoint \tpath{s,t}s.
    Putting this together with \autoref{cor:cuts_XP}, we get that testing whether $\tp_{{\cal G}}(s,t) \geq 2$ can be done in time $O(n\tau)$ (as it can be solved for $h=1$ insteadh of $h=2$).
\end{proof}

Now, we use \autoref{lem:tpc-2-iff-tp-at-least-2} to \emph{find}~2 \tvertex-disjoint \tpath{s,t}s, if they exist. The idea is that we can delete a temporal edge of $\mathcal{G}$ while maintaining at least 2 disjoint \tpath{s,t}s in the resulting temporal graph. We first prove two useful lemmas, and for this we need a new definition.
We say that $\mathcal{G}$ is \emph{$s,t$-minimal} if $\tp_{\mathcal{G}}(s,t) = 2$, and $\tp_{\mathcal{G} - (e,i)}(s,t) = 1$ for every $(e,i) \in E^T(\mathcal{G})$.

\begin{lemma}
    Let ${\cal G} = (G,\lambda)$ be a temporal (directed) graph on $n$ vertices, $m$ edges and lifetime $\tau$, and consider non-adjacent vertices $s,t\in V(G)$. We can, in time $O(mn\tau^2)$, either output an $s,t$-minimal temporal (directed) graph ${\cal G}'\subseteq G$, or conclude that $tp(s,t)\le 1$.
\end{lemma}
\begin{proof}
    By \autoref{lem:tpc-2-iff-tp-at-least-2}, we can decide whether $tp(s,t)\le 1$ in time $O(n\tau)$, so from now on we suppose that this is not the case, i.e., that $tp(s,t)\ge 2$.
    Now, consider an order of the temporal edges of ${\cal G}$, $e_1,\ldots, e_{m_T}$, and let ${\cal G}'$ be initially equal to ${\cal G}$. For each $i\in [m_T]$, in this order, test (in time $O(n\tau)$ by \autoref{lem:tpc-2-iff-tp-at-least-2}) whether $tp_{{\cal G}'-e_i}(s,t)\ge 2$. If so, then remove $e_i$ from ${\cal G}'$; otherwise continue to the next temporal edge. We claim that, at the end, ${\cal G}'$ is $s,t$-minimal. We get running time $O(mn\tau^2)$ since $m_T\le m\tau$.

    Because we do not remove a temporal edge if the removal decreases $tp_{{\cal G}'}(s,t)$ to~1, we know that $tp_{{\cal G}'}(s,t)\ge 2$. Additionally, suppose that $tp_{{\cal G}'}(s,t)> 2$ and let $e_i$ be a temporal edge contained in any among 3 \tvertex-disjoint \tpath{s,t}s. Then $tp_{{\cal G}'-e_i}(s,t)\ge 2$ and $e_i$ should have been removed in its iteration. Hence $tp_{{\cal G}'}(s,t)= 2$. Observe that a similar argument can be applied in case $tp_{{\cal G}'-e_i}(s,t)\ge 2$ for some $e_i\in E^T({\cal G}')$. As $tp_{{\cal G}'}(s,t)$ decreases by at most one with a temporal edge removal, we get $tp_{{\cal G}'-e_i}(s,t) = 1$ for every $e_i\in E^T({\cal G}')$, and it follows that ${\cal G}'$ is $s,t$-minimal, as we wanted to prove.
\end{proof}

It remains to prove that~2 \tvertex-disjoint paths within an $s,t$-minimal temporal graph can be found in polynomial time.

\begin{lemma}\label{lemma:finding_paths_in_minimal_temporal_graph}
    Let $\mathcal{G} = (G, E)$ be a temporal (directed) graph, and consider $s,t\in V(G)$. If ${\cal G}$ is $s,t$-minimal,
    then we can find 2 \tvertex-disjoint \tpath{s,t}s in $\mathcal{G}$ in $\mathcal{O}(n)$ time, where $n = |V(G)|$.
\end{lemma}
\begin{proof}
    Since $\tp_{\mathcal{G}}(s,t) = 2$ and $\tp_{\mathcal{G}-e}(s,t) = 1$, for every $e\in E^T({\cal G})$, we get that
    every temporal edge of $\mathcal{G}$ is contained in some \tpath{s,t}.
    Since every such path contains exactly 2 temporal edges incident to each internal vertex, as well as exactly 1 temporal edge incident to $s$ and 1 to $t$, we also conclude that $d^T(s) = d^T(t) = 2$ and that $d^T(v) \in  \{0,2,4\}$ for every $v \in V(G) \setminus \{s,t\}$. As a consequence, we get that $\lvert E^T({\cal G})\rvert \le 4n$; as we will see shortly, our algorithm simply does a search on $G$ twice, and this is why we get running time $O(n)$.

    We iteratively construct a partial \tpath{s,t}, denoted by $P_i$, starting in $s$, while ensuring the following property.

    \medskip
    \noindent \textbf{Property ($\star$):}
    There exist 2 \tvertex-disjoint \tpath{s,t}s in ${\cal G}$, $P$ and $Q$, such that $P_i\subseteq P$.

    \medskip

    We begin with $V(P_0) = \{s\}$ and $v_0 = s$. Then, for $i \geq 0$, we construct $P_{i+1}$ from $P_i$ by appending a temporal edge $(v_iv_{i+1},j_i)$ while ensuring Property $(\star)$.
    Since $d^T(s) = 2$, we can safely append to $P_0$ any temporal edge leaving $s$ since $\mathcal{G}$ is $s,t$-minimal.
    Assume now that $i \geq 1$ steps have been performed, and let $v_i$ be the last vertex of $P_i$.
    Suppose first that $v_i\neq t$.
    If $d^T(v_i) = 2$, then there is only one temporal edge $(e,j)$ incident to $v_i$ that is not used by $P_i$.
    By Property $(\star)$ we know that, since $P_i\subseteq P$ and $v_i\neq t$, we get that $P$ must also contain $(e,j)$; hence we simply add $(e,j)$ to $P_i$ to construct $P_{i+1}$.
    Clearly Property $(\star)$ holds for $P_{i+1}$.

    Now, suppose that $d^T(v_i) = 4$, in which case there are exactly three temporal edges adjacent to $v_i$ not being used by $P_i$.
    Let $e_i = (v_{i-1}v_i,j_i)$ be the temporal edge of $P_i$ incident to $v_i$, and let $e_{i+1} = (v_iv_{i+1},j_{i+1}) \in \delta^T(v_i)\setminus \{e_i\}$ be a temporal edge incident in $v_i$ in a timestep $j_{i+1}\ge j_i$ chosen as to minimize $j_{i+1}$.
    We construct $P_{i+1}$ by appending $e_{i+1}$ to $P_i$.
    Now, consider the paths $P,Q$ ensured by Property $(\star)$. Observe that if it holds that $e_{i+1}\in E^T(P)$, then we are done. This is what we prove next.

    Let $(f_1, \ell_1), (f_2, \ell_2)$ be the temporal edges of $\delta^T(v_i) \setminus \{e_i,e_{i+1}\}$.
    Because $P_i\subseteq P$ and $P,Q$ are \tvertex-disjoint, and since ${\cal G}$ is $s,t$-minimal, we know that exactly 2 temporal edges of $\delta^T(v_i)$ are in $P$ (including $e_i$), and exactly 2 are in $Q$. Suppose by contradiction that $(f_1,\ell_1)\in E^T(P)$. Then, since $(f_1,\ell_1)$ occurs after $e_i$ in $P$, we get that $\ell_1\ge j_i$. Additionally, by the choice of $e_{i+1}$, we must have $\ell_1 = j_{i+1}$. This is a contradiction as in this case we have that $e_{i+1}$ must be in $Q$, and $(v_i,\ell_1)\in V^T(P)\cap V^T(Q)$.

    Finally, if $v_i = t$, then observe that we can apply the same procedure to ${\cal G}-E^T(P)$ in order to find the second path $Q$ insured by Property $(\star)$.
\end{proof}

\autoref{thm:finding-2-temporal-paths-poly-algorithm} directly follows from the above lemmas.

\begin{theorem}\label{thm:finding-2-temporal-paths-poly-algorithm}
    Let ${\cal G} = (G, \lambda)$ be a temporal (directed) graph with lifetime $\tau$ and $s,t \in V(G)$.
    We can either find 2~\tvertex-disjoint \tpath{s,t}s in ${\cal G}$, if they exist, or output $\tp{(s,t)} \leq 1$ otherwise. The running time is $O(mn\tau^2)$, where $n = \lvert V(G) \rvert$ and $m = \vert E(G) \rvert$.
\end{theorem}

It is worth observing that, consistently with \autoref{thm:paths_negative_directed} which tells us that \ktpaths is $\NP$-complete on directed graphs even if $k=3$, the algorithm previously described cannot be generalized for bigger values of $k$. The main reason is that we do not have an oracle that tells us whether $\tp_{\mathcal{G}'}(s,t) \geq k$ in polynomial time unless $\P=\NP$. 


\section{Conclusion}

Given a temporal graph ${\cal G} = (G,\lambda)$ with lifetime $\tau$ and non-adjacent vertices $s,t\in V(G)$, in this paper  we have thoroughly investigated problems related to connectivity and robustness of a given pair of vertices. More specifically, we have worked on disjoint paths and cut problems whose definitions are based on vertex and \tvertex-disjointness, $G$ being undirected or directed, and temporal paths being strict or non-strict paths. Ours and previous results can be found in \Cref{tab:summarywalks_nonstrict,tab:summarywalks_strict}. We encourage the reader to carrefully compare the two tables and find interesting and challenging open question. Here, we propose some that derived directly from such comparison, in addition to some open questions that are stated within the text.

From what is commented in the introduction, there is a case that is left open in~\cite{KKK.00} when they answer a question proposed in~\cite{B.96}:

\begin{question}
    In the non-strict context, given a temporal graph $\mathcal{G} = (G,\lambda)$ and non-adjacent vertices $s,t$, can we decide whether there exist 2 vertex-disjoint \tpath{s,t}s in polynomial time?
\end{question}

A related question that can be spotted when comparing the two tables is the following:

\begin{question}
    In the strict context, given a temporal directed $\mathcal{G} = (G,\lambda)$ and non-adjacent vertices $s,t$, can we decide whether there exist 3 \tvertex-disjoint \tpath{s,t}s in polynomial time? What about if we fix also $\tau$?
\end{question}

As we commented in the introduction, the example in \autoref{fig:tvertexpaths_counter} can be used to obtain a temporal graph $\mathcal{G} = (G,\lambda)$ within the strict context for which $\tp(s,t) = 2 < \tpc(s,t)=3$. We already know that this is best possible because of \autoref{thm:menger}. However, the obtained graph has lifetime~6. Therefore, we ask:

\begin{question}
    In the strict context, what is the minimum value $\tau$ for which there exist a temporal graph $\mathcal{G} = (G,\lambda)$ and non-adjacent vertices $s,t$ such that $\tp(s,t) = 2 < \tpc(s,t)=3$?
\end{question}

Finally, in~\cite{IS.24_JCSS}, the authors point out in their final remarks the fact that Menger's Theorem does not hold for the temporal edge-disjointt case. There, they are insterested in what is called \emph{Mengerian graphs}, a notion introduced in the seminal paper~\cite{KKK.00} and generalized by the authors. Here, we propose a thorough investigation of edge based Menger's Theorem.

\begin{question}
    How would \Cref{tab:summarywalks_nonstrict,tab:summarywalks_strict} look like if we consider edge/temporal edge-disjointt walks/paths?

\end{question}

\bibliography{references.bib}

\clearpage

\appendix

\section{Understanding the work of Mertzios et al.~\cite{MMS.19}.}\label{app:mertzios-paths-and-walks}
Mertzios et al.~\cite{MMS.19} study the problem of temporal vertex-disjoint paths in temporal graphs.
Adopting our notation, they define a \emph{journey} $J$ of a temporal graph $\mathcal{G} = (G, \lambda)$ as a path of the static graph $G$ with edges $(e_1, \ldots, e_k)$ and labels $\ell_i \in \lambda(e_i)$, for $i \in [k]$, such that $\ell_1 < \ell_2 < \cdots < \ell_k$.
Thus journeys are equivalent to our definition of (strict) temporal paths, not walks, if one understands paths as we define them.
We acknowledge that the notation is not standardized: in some works, the term \emph{path} allows repetition of vertices, while the term \emph{simple path} does not.
With the definition of journeys at hand, the authors state an analogous of Menger's Theorem for journeys and temporal vertex cuts.

\begin{proposition}[Mertzios et al.~\cite{MMS.19}]\label{prop:temporal-journeys-menger}
    Let $\mathcal{G} = (G, \lambda)$ be a temporal graph and $s,t \in V(G)$.
    The maximum number of out-disjoint journeys from $s$ to $t$ is equal to the minimum number of vertex departure times needed to separate $s$ from $v$.
\end{proposition}
In other words, with the understanding that journeys are temporal paths, \autoref{prop:temporal-journeys-menger} implies that $\tp_{\mathcal{G}}(s,t) = \tpc_{\mathcal{G}}(s,t)$ and contradicts \autoref{thm:menger}.
For clarity, we argue that \autoref{prop:temporal-journeys-menger} holds for temporal walks only.

The proof of \autoref{prop:temporal-journeys-menger} relies on the static expansion $D$ (see, for example, the proof of \autoref{thm:tvertex_vwalks}, \cite{CLMS.21,Skutella2009}, or \cite[Definition 2]{MMS.19} for the definition) of the given temporal graph $\mathcal{G}$.
They begin by observing the maximum number of disjoint paths from $s$ to $t$ in the temporal expansion, and then associate these paths to journeys in the temporal graph.
This reliance on temporal expansions implies that \autoref{prop:temporal-journeys-menger} is true only for temporal \emph{walks}, and {\sl not} for temporal \emph{paths}, since paths in the temporal expansion are not guaranteed to yield temporal paths in the temporal graph.
See, for example, \autoref{fig:journeys-vs-paths-static-expansion}.
The blue path on the temporal expansion (on the right side), needs to take the detour from $u$ to $x$ then back in order to allow room for the red path through $u$.
When translated back to the temporal graph (on the left), the blue path is associated with a temporal walk.
In this particular example, there are two \tvertex-disjoint \twalk{s,t}s in the temporal graph, but the two \tvertex \tpath{s,t}s intersect.

This justifies the existence of \autoref{thm:menger}: when restricted to temporal paths only, we show that an analogous to Menger's Theorem holds if and only if the maximum number of temporal vertex-disjoint paths is $1$.

\begin{figure}][h]
    \centering
    \begin{tikzpicture}[
            box/.style={rectangle,draw,minimum width=1cm,minimum height=1cm},
        ]
        
        \begin{scope}]
            \node[draw, circle, fill=black, scale=.6, label=-90:{$s$}] (s) at (0,0) {};
            \node[draw, circle, fill=black, scale=.6, label=90:{$t$}] (t) at ($(s) + (0,4)$) {};

            \node[draw, circle, fill=black, scale=.6, label=180:{$u$}] (u) at ($(s) + (0,2)$) {};
            \node[draw, circle, fill=black, scale=.6, label=9:{$v$}] (v) at ($(s) + (2,2)$) {};


            \draw[->, >=latex, goodblue] (s) to [bend left = 30] node[midway, xshift=-4pt] {$1$} (u);
            \draw[->, >=latex, goodred] (s) to [bend right = 30] node[midway, xshift=-4pt] {$3$} (u);

            \draw[->, >=latex, goodblue] (u) to [bend left = 30] node[midway, above] {$2$} (v);
            \draw[->, >=latex, goodblue] (v) to [bend left = 30] node[midway, above] {$5$} (u);
            \draw[->, >=latex, goodblue] (u) to [bend left = 30] node[midway, xshift=-4pt] {$6$} (t);
            \draw[->, >=latex, goodred] (u) to [bend right = 30] node[midway, xshift=-4pt] {$4$} (t);
            



        \end{scope}%

        \begin{scope}[xshift=7cm, yshift=-2cm]
            \foreach \i/\val in {6/6,5/5,4/4,3/3,2/2,1/1} {
                    \node[draw, circle, fill=black, scale=.6] (L\i) at (0, \i) {};
                    \node (n\i) at ($(L\i) + (-1, 0)$) {\large$\i$};
                }
            \node (tau) at ($(n6) + (0,0.5)$) {\Large$\tau$};
            \foreach \i/\val in {6/6,5/5,4/4,3/3,2/2,1/1} {
                    \node[draw, circle, fill=black, scale=.6] (R\i) at (2, \i) {};
                }


            \node[draw, rectangle, fit=(L1)(L6), label=93:$u$] {};
            \node[draw, rectangle, fit=(R1)(R6), label=87:$v$] {};


            \node[draw, circle, fill=black, scale =.6, label=-90:{$s$}] (s) at ($(L1) + (1, -1)$) {};
            \node[draw, circle, fill=black, scale =.6, label=90:{$t$}] (t) at ($(L6) + (1, 1)$) {};

            \draw[->, >=latex, goodblue] (s) to (L1) to (L2) to (R2) to (R3) to (R4) to (R5) to (L5) to (L6) to (t);

            \draw[->, >=latex, goodred] (s) to [out = 90, in = -45] (L3) to (L4) to [out = 0, in = -90] (t);
        \end{scope}
        
    \end{tikzpicture}
    \caption{Example of paths in the static expansion (on the right) and temporal walks (on the left). In the static expansion, other edges between occurrences of the same vertex are ommited. For simnplicity, the ``expansions'' of $s$ and $t$ are also ommited. Notice that all temporal walks are strict.}
    \label{fig:journeys-vs-paths-static-expansion}
\end{figure}
\end{document}